\documentclass[traditabstract]{aa} 
\usepackage{graphicx}
\usepackage{txfonts}
\usepackage{natbib}
\usepackage{epsfig}
\usepackage{deluxetable}
\usepackage{longtable}

\def\simgt{\lower.5ex\hbox{\gtsima}}

\newcommand{\s}{{\it Spitzer}}
\newcommand{\h}{{\it Herschel}}

\newcommand{\md}{$M_{\rm dust}$}

\newcommand{\lsol}{L$_{\odot}$}
\newcommand{\lir}{$L_{\rm IR}$}

\newcommand{\td}{$T_{\rm d}$}

\newcommand{\lmid}{$L_{\rm 8}$}
\title{Mid- to far infrared properties of star-forming galaxies and active galactic nuclei
\thanks{Herschel is an ESA space observatory with science instruments provided
  by European-led Principal Investigator consortia and with important participation from NASA.}}

\author{Georgios E. Magdis\inst{1}
\and D.~Rigopoulou \inst{1,2}
\and G. Helou \inst{3}
\and D. Farrah \inst{4}
\and P. Hurley \inst{6}
\and A. Alonso-Herrero \inst{6,7}
\and J. Bock \inst{8,9}
\and D. Burgarella \inst{10}
\and S. Chapman \inst{11}
\and V. Charmandaris \inst{12,13,14}
\and A. Cooray \inst{15}
\and Y. Sophia Dai \inst{16}
\and D. Dale \inst{17}
\and D. Elbaz \inst{18} 
\and A. Feltre \inst{19,20}
\and E. Hatziminaoglou \inst{20}
\and J-S. Huang \inst{16} 
\and G. Morrison \inst{21,22}
\and S. Oliver \inst{6}
\and M. Page \inst{23}
\and D. Scott \inst{24}
\and Y. Shi \inst{25}
}
\institute{Department of Physics, University of Oxford, Keble Road, Oxford OX1 3RH, UK \\
\email:{georgios.magdis@astro.ox.ac.uk}
\and Space Science \& Technology Department, Rutherford Appleton Laboratory, Chilton, Didcot, Oxfordshire OX11 0QX, UK
\and Infrared Processing and Analysis Center, California Institute of Technology, Pasadena, CA 91125, USA
\and Virginia Polytechnic Institute and State University Department of Physics, MC 0435, 910 Drillfield Drive, Blacksburg VA 24061, USA
\and Astronomy Centre, Department of Physics and Astronomy, University of Sussex, Falmer, Brighton BN1 9QH
\and Instituto de Fisica de Cantabria, CSIC-UC, 39005 Santander, Spain
\and Augusto G. Linares Senior Research Fellow
\and California Institute of Technology, 1200 E. California Boulevard, Pasadena, CA 91125, USA,
\and Jet Propulsion Laboratory, 4800 Oak Grove Drive, Pasadena, CA 91109, USA
\and Aix-Marseille Universit\'e, CNRS, LAM (Laboratoire d'Astrophysique de Marseille)
\and Dalhousie University, Halifax, Nova Scotia B3H 4R2, Canada
\and Department of Physics and ITCP, University of Crete, GR-71003 Heraklion, Greece 
\and IESL/Foundation for Research and Technology - Hellas, GR-71110, Heraklion, Greece 
\and Chercheur Associ\'e, Observatoire de Paris, F-75014, Paris, France
\and Department of Physics \& Astronomy, University of California, Irvine, CA 92697, USA
\and Harvard-Smithsonian Center for Astrophysics, 60 Garden Street, MS65, Cambridge, MA 02138, USA
\and Department of Physics and Astronomy, University of Wyoming, Laramie, WY 82071, USA
\and Laboratoire AIM-Paris-Saclay, CEA/DSM/Irfu, CNRS, Universit\'e Paris Diderot, Saclay,  Gif-sur-Yvette, France
\and ESO, Karl-Schwarzschild-Str. 2 85748 Garching bei Munchen, Germany
\and Dipartimento di Fisica e Astronomia, Vicolo Osservatorio 2 I-35122 Padova, Italy  
\and Institute for Astronomy, University of Hawaii, Honolulu, HI 968226, USA
\and Canada-France-Hawaii Telescope, Kamuela, HI 96743, USA
\and Mullard Space Science Laboratory, University College London, Holmbury St Mary Dorking, Surrey RH5 6NT
\and Department of Physics \& Astronomy, University of British Columbia, 6224 Agricultural Road, Vancouver, BC V6T 1Z1, Canada
\and School of Astronomy And Space Science, Nanjing University, Jiangsu 210093, China
}
\begin{document}
\abstract{We study the mid- to far-IR properties of a 24$\,\mu$m-selected flux-limited sample ($S_{\rm 24}$ $>$ 5$\,$mJy) of 154 intermediate redshift ($\langle z \rangle \sim 0.15$), infrared luminous  galaxies, drawn from the 5 Milli-Jansky Unbiased Spitzer Extragalactic Survey. By combining existing mid-IR spectroscopy and new \h\ SPIRE submm photometry from the Herschel Multi-tiered Extragalactic Survey, we derived robust total infrared luminosity (\lir) and dust mass (\md) estimates and infered the relative contribution of the AGN to the infrared energy budget of the sources. We found that the total (8$-$1000$\,\mu$m) infrared emission of galaxies with weak 6.2$\,\mu$m PAH emission ($EW_{\rm 6.2}$ $\leq$ 0.2$\,\mu$m) is dominated by AGN activity, while for galaxies with $EW_{\rm 6.2}$ $>$ 0.2$\,\mu$m more than 50\% of the \lir\ arises from star formation. We also found that for galaxies detected in the 250-500$\,\mu$m \h\ bands an AGN has a statistically insignificant effect on the temperature of the cold dust and the far-IR colours of the host galaxy, which are primarily shaped by star formation activity. For star-forming galaxies we reveal an anti-correlation between the \lir-to-rest-frame 8$\,\mu$m luminosity ratio, IR8 $\equiv$ \lir/\lmid\, and the strength of PAH features. We found that this anti-correlation is primarily driven by variations in the PAHs emission, and not by  variations  in the 5$-$15$\,\mu$m mid-IR continuum emission. Using the [\ion{Ne}{III}]/[\ion{Ne}{II}] line ratio as a tracer of the hardness of the radiation field, we confirm that galaxies with harder radiation fields tend to exhibit weaker PAH features, and found that they have higher IR8 values and higher dust-mass-weighted luminosities (\lir/\md), the latter being a proxy for the dust temperature (\td). We argue that these trends originate either from variations in the environment of the star-forming regions or are caused by variations in the age of the starburst. Finally, we provide scaling relations that will allow estimating \lir, based on single-band observations with  the mid-infrared instrument,  on board the upcoming  James Webb space telescope.}

\titlerunning{Mid to Far IR properties of star-forming galaxies and AGN}
\authorrunning{G. E. Magdis et al.}
\maketitle
\section{Introduction}
One of the major advances in our understanding of galaxy evolution was the discovery of the cosmic infrared background (CIB), first detected by Puget et al. (1996), which led to the realisation that half of the energy produced by star formation and accretion activity throughout the history of the Universe is emitted via the infrared (8$-$1000$\,\mu$m) part of the spectrum (Dole et al. 2006). Because in the local Universe the infrared output of galaxies is only about a third of the emission at optical wavelengths (e.g., Soifer \& Neugebauer 1991), this implies a strong evolution of infrared galaxy populations, with an enhanced far-IR output in the past, to account for the total measured CIB. 

With the advent of infrared space telescopes, such as the Infrared Space Observatory (Kessler et al. 2996, see Genzel \& Cesarsky 2000, for a summary), the Spitzer Space Telescope (\s, Werner et al. 2004), and more recently the Herschel Space Observatory (\h, Pilbratt et al. 2010) we have been able to resolve a considerable fraction of the CIB (e.g. $\sim$74\% at 160$\,\mu$m, Berta et al. 2011)  into individual sources and construct and characterise large samples of infrared galaxies (for a review see Lagache et al. 2005). These infrared campaigns have revealed that the number density of luminous infrared galaxies (LIRGs \lir\ $>$ 10$^{11}$ \lsol) and ultra luminous infrared galaxies (ULIRGs \lir\ $>$ 10$^{12}$ \lsol), which emit the bulk of their energy in the infrared, increases by three orders of magnitude as we look back in time and they also dominate the star formation density of the Universe by $z\sim1$ (e.g. Le Floc'h et al. 2005). 

Different parts of the IR spectrum of galaxies are dominated by different physical processes. The mid-IR (5$-$25$\,\mu$m) emission  is dominated by warm dust emission, which originates from the small dust grains that are heated by energetic photons produced by young stars or through AGN accretion activity.  Superimposed on this continuum emission is a large set of broad emission line features, centred at 3.3, 6.2, 7.7, 8.6, 11.2, and 12.7$\,\mu$m,  which are thought to originate from polycyclic aromatic hydrocarbons (PAHs; Puget et al. 1985; Allamandola et al. 1989). PAH emission predominantly originates from  photodissociation regions (PDRs) that are illuminated by UV-bright stars  and can contribute up to 10\% of the total infrared luminosity (\lir) in star-forming galaxies (e.g. Smith et al. 2007). Previous \s\ studies using the Infrared Spectrograph (IRS, Houck et al. 2004) have shown that PAH emission varies considerably between star-forming and AGN-dominated galaxies and as a function of the metal enrichment of the interstellar medium (ISM) of a galaxy. In particular, PAHs are found to be prominent in star-forming galaxies or weak AGNs both locally (Peeters et al. 2002, 2004; Weedman et al. 2005; Armus et al. 2007) and at high$-z$ (Pope et al. 2008; Huang et al. 2009; Desai et al. 2009), while they tend to be weaker or even absent in galaxies dominated by an AGN (e.g. Kirkpatrick et al. 2012; Weedman et al. 2006). Hence, the strength and the relative ratio of the PAHs, along with a plethora of other atomic fine structure lines (e.g., Ne, S)  and absorption by amorphous silicates centred at 9.7$\,\mu$m, offer a unique diagnostic tool characterising  the dominant mechanism that powers the IR emission of the galaxies (e.g., Laurent et al. 2000; Lutz et al. 1996; Genzel et al. 1998; Peeters et al. 2004; Spoon et al. 2007). 

Mid-IR spectroscopic studies of local (Rigopoulou et al. 1999; Farrah et al. 2007; Desai et al. 2007) and distant (e.g., Yan et al. 2007; Dasyra et al. 2009; Hernan-Caballero et al. 2009) LIRGs and ULIRGs have revealed that while both AGN and star formation activity is present, their IR emission is mostly powered by star-formation. However, to properly measure the bolometric output of the galaxies there is an imperative need for far-IR (50-350$\,\mu$m) observations that trace the peak of the spectral energy distribution (SED) of star-forming galaxies. Furthermore, while recent studies of high$-z$ galaxies have come to suggest that the far-IR emission is primarily shaped by star formation activity (Hatziminaoglou et al. 2010; Kirkpatrick et al. 2012; Feltre et al. 2013), it is still unclear whether and how an AGN could impact the far-IR colours and the dust temperature of the large grains in the ISM of a galaxy (e.g. Haas et al. 2003; Netzer et al. 2007). 

With the advent of \h\ we have now gained access to the far-IR  part of the spectrum of the galaxies up to $z \sim 2$, and it is now possible to bridge the warm and cold dust emission and reveal the underlying heating mechanisms. In this direction, it has recently been demonstrated that  for the majority of star-forming galaxies up to $z \sim 2$, the ratio of total (\lir) to the mid-IR luminosity, as traced by the rest-frame 8$\,\mu$m emission (\lmid), obeys an almost linear relation, suggesting a very uniform mid-to-far-IR SED shape for star-forming galaxies through cosmic time (Elbaz et al.\ 2011). However, outliers to this relation do exist, with a small fraction of star-forming galaxies exhibiting an enhanced  IR8 $\equiv$ \lir/\lmid\ value. While this excess in IR8 is found to anti-correlate  with the IR surface brightness and hence with the projected star formation density (Elbaz et al. 2011), the true nature of these sources, the origin and the scatter of the observed \lir$-$\lmid\ relation among galaxies that fall in this ``IR main sequence'', and the connection of the mid- to the far-IR properties of the galaxies are yet to be fully understood. 

Another key question that still remains open is whether the characteristics of the PAH features can be used to infer information about the  star formation properties of the galaxies. In principle, PAHs could serve as a good tracer of star formation activity, since they are stochastically heated mainly by UV photons produced by stars (PAHs can also be excited by visual photons, although the excitation is dominated by UV photons; e.g., Uchida et al. 1998; Li \& Draine 2002). Under the assumption of fixed emission and absorption properties and fixed PAH abundance, the PAH emission is a measure of the amount of photons available between 6 and 13.6 eV and hence of star formation. This line of reasoning is supported by various studies that have demonstrated an almost linear correlation between PAH emission and \lir\ (e.g. Soifer et al. 2002; Peeters et al. 2004; Lutz et al. 2008), the latter thought to be an excellent tracer of star-formation rate for dusty circumnuclear starbursts (Kennicutt 1998). However, the  $L_{\rm PAH}$/\lir\ ratio is known to vary as a function of the environment in which star formation takes place (e.g., Peeters et al. 2004) but also as a function of AGN activity (e.g. Siebenmorgen et al. 2004). This also applies to integrated $L_{\rm PAH}$ and \lir\ measurements over the whole galaxy. In particular, the $L_{\rm PAH}$/\lir\ ratio can vary up to a factor 10 between sources where the environment of massive star formation resembles PDRs (e.g. M82, Carral et al. 1994) and that of embedded star formation (e.g. Arp220) that is more similar to the properties of compact \ion{H}{II} regions (Rigopoulou et al. 1999). Interestingly, a similar variation in the strength of far-IR fine structure lines is observed among different star formation environments (e.g. Gracia-Carpio et al. 2011). It could therefore be suggested that the integrated PAH properties of distant galaxies, where the individual star-forming regions remain unresolved, can carry crucial information about the averaged star formation activity on a galaxy-wide scale.

To address these questions, in this paper we exploit a 24$\,\mu$m-flux-limited sample of intermediate redshift mid-IR-selected  galaxies that benefit from  high-quality mid-IR IRS spectra obtained as part of the 5 Milli-Jansky Unbiased Spitzer Extragalactic Survey (5MUSES). By combining existing mid- and far-IR data  with new submm (250$-$500$\,\mu$m) \h-SPIRE observations drawn from the HerMES project, we attempt a full characterisation of their IR properties and study the variation of the PAH features among star-forming galaxies. A key aspect of our sample is that it traces the epoch of the steep increase of the infrared luminosity density ($0.0 \leq z \leq 1.0$),  providing a snapshot of the evolution of the properties of  star-forming galaxies from the present day to the peak of the star formation activity in the Universe ($z \sim$ 1$-$2). In Section 2 we provide a description of the sample, present ancillary IRS and newly obtained \h\ data  and classify the dominant powering source of the galaxies. In Section 3 we derive the far-IR properties of the sample, while in Section 4 we investigate the impact of the AGN activity on the far-IR properties of the galaxies and examine the variation of the strength of PAH features as a function of IR8. In Section 6 we provide a discussion motivated by our results, and finally in Section 7 we 
summarise our findings. Throughout the paper we adopt $\Omega_{\rm m}$ = 0.3, $H_{\rm 0}$ = 71 km s$^{-1}$ Mpc$^{-1}$ and $\Omega_{\rm \Lambda}$ = 0.7.
 
\section{Sample selection and observations}
5MUSES is a 24$\,\mu$m flux-limited (5$\,$mJy $< S_{\rm 24}$ $<$ 100$\,$mJy), mid-IR spectroscopic survey of 330 galaxies selected from the SWIRE fields (Lonsdale et al. 2003), including Elais-N1, Elais-N2, Lockman Hole, and XMM-LSS, in addition to the \s\ Extragalactic First Look Survey (XFLS) field (Fadda et al. 2006). Out of this sample we focus on the 280 sources presented by Wu et al. (2010), for which we have secure spectroscopic redshifts. This simple selection criterion of the 5MUSES survey provides an intermediate-redshift sample ($\langle z \rangle $ = 0.144) they bridges the gap between the bright, nearby star-forming galaxies (e.g. Kennicutt et al. 2003; Smith et al. 2007; Dale et al. 2009; Pereira-Santaella et al. 2010; Diaz-Santos et al. 2010), local (U)LIRGs (e.g. Armus et al. 2007; Desai et al. 2007; Imanishi et al. 2007; Farrah et al. 2007;  Veilleux et al. 2009), and the more distant sources that have been followed up with IRS spectroscopy (Houck et al. 2005; Sajina et al. 2007; Yan et al. 2007; Farrah et al. 2008;  Pope et al. 2008; Desai et al. 2009; Men{\'e}ndez-Delmestre et al. 2009; Sargsyan et al. 2011 ). Low-resolution mid-IR spectra ($R = 64-128$) of all galaxies in 5MUSES have been obtained with the short-low (SL: $5.5-14.5\,\mu$m) and long-low (LL: $14-35\,\mu$m) modules of the IRS using the staring mode observations. A full description of the IRS observations and data reduction are presented in Wu et al. (2010). In brief, PAH luminosities and equivalent width (EW) were measured using 
the PAHFIT software (Smith et al. 2007) as well as a spline-fitting method. In the former, the PAH features are fit with Drude profiles, which have extended wings 
that account for a significant fraction of the underlying plateau (Smith et al. 2007), while in the  spline or apparent PAH EW method, a local continuum 
under the emission features is defined by fitting a spline function to selected continuum points. While the PAHFIT method is known to give higher integrated PAH 
fluxes and EWs due to the lower continuum adopted than the spline method, the two methods  provide consistent results on trends (Smith et al. 2007; Galliano et al. 2008). Following Wu et al. (2010), we adopt PAH EWs as measured by the  spline-fitting method, fixing the rest wavelength continuum pivots as in Peeters et al. (2002), and 6.2-, 7.7-, and 11.3$\,\mu$m PAH-integrated fluxes as derived from PAHFIT. This was done  to be able to compare our measurements with other samples reported in the literature, because the majority of these use the spline method when reporting EW values. Finally, in addition to the 24$\,\mu$m photometry and IRS spectroscopy, the whole sample benefits from 70-160$\,\mu$m MIPS and  IRAC 3.6$-$8.0\,$\mu$m observations, with 90\% and  54\% of the sources being detected at 70 and 160$\,\mu$m, respectively.

\subsection{\h\ observations}
We used \h\ SPIRE (Griffin et al. 2010) observations of the fields  ELAIS-N1, Lockman Hole, XMM, and XFLS, obtained as part of the Herschel Multi-Tiered Extragalactic Survey{\footnote{For more information about the HerMES programme visit hermes.sussex.ac.uk}} 
(HerMES; Oliver et al. 2010, 2012). For these sources, we employed the photometric catalogues at 250, 350, and 500$\,\mu$m that were produced for each field by using a prior source extraction, guided by the position of known 24$\,\mu$m sources. An extensive description of the cross-identification prior source 
extraction (XID) method is given in Roseboom et al. (2010, 2012). The main advantage of this method is that reliable fluxes can be extracted close to 
the formal $\approx$ 4 $-$ 5\,mJy SPIRE confusion noise (Nguyen et al. 2010) by estimating the flux contributions from nearby sources within one 
beam. The 24$\,\mu$m prior positional information reduces the impact of confusion noise and so the approximate 3$\sigma$ limit for the SPIRE catalogue at 
250$\,\mu$m  is $\approx$ 9$-$15$\,$mJy.  The drawback of this technique is that the resulting catalogues could be missing sources 
without a 24$\,\mu$m counterpart, that is, 24$\,\mu$m drop-outs (e.g. Magdis et al. 2011). However, since all galaxies in the 5MUSES sample have a bright 24$\,\mu$m counterpart, we are not affected by this caveat. 

Out of the 280 sources in the 5MUSES sample presented by Wu et al. (2010), 188 galaxies are covered by HerMES observations. After applying a  flux cut limit of 15$\,$mJy to all three bands of our SPIRE photometric catalogues, 154 sources are  detected at 250$\,\mu$m  at a 3$\sigma$ significance level, 108 at 350$\,\mu$m, and 50 at 500$\,\mu$m. All sources detected at 350 and 500$\,\mu$m are also detected at 250$\,\mu$m. To assess the robustness of the detections we also preformed a visual inspection of the sources in the \h\ maps. The SPIRE photometry of the final sample, that is, sources with at least one detection at one of the \h\  bands, is presented in Table 1. The spectroscopic redshifts of the sources are drawn from Wu et al. (2010) and the median redshift of the sample considered here is  $\langle z \rangle = 0.157 $. Finally,  a K$-$S test reveals that the $S_{\rm 250}$ values  of the  whole population of galaxies in our fields with $S_{\rm 24}$ $>$ 5$\,$mJy and that of the 5MUSES sample are drawn from the same distribution. While this was expected based on the simple selection criteria of the 5MUSES samples ($S_{\rm 24}$ $>$ 5$\,$mJy), it also suggests that it is representative of the full HerMES population at this 24$\,\mu$m flux limit.

\begin{figure*}
\centering
\includegraphics[scale=0.39]{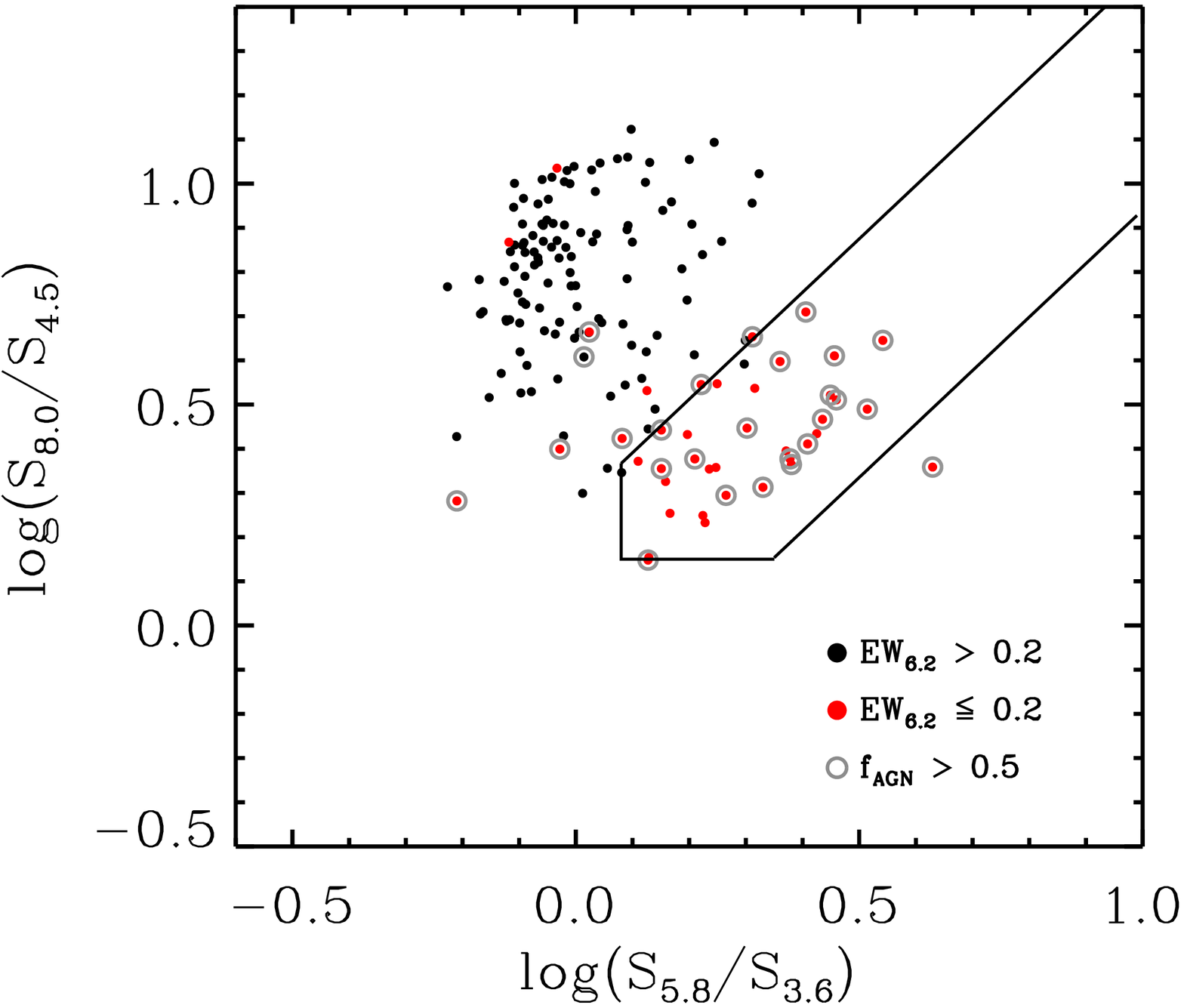}
\includegraphics[scale=0.39]{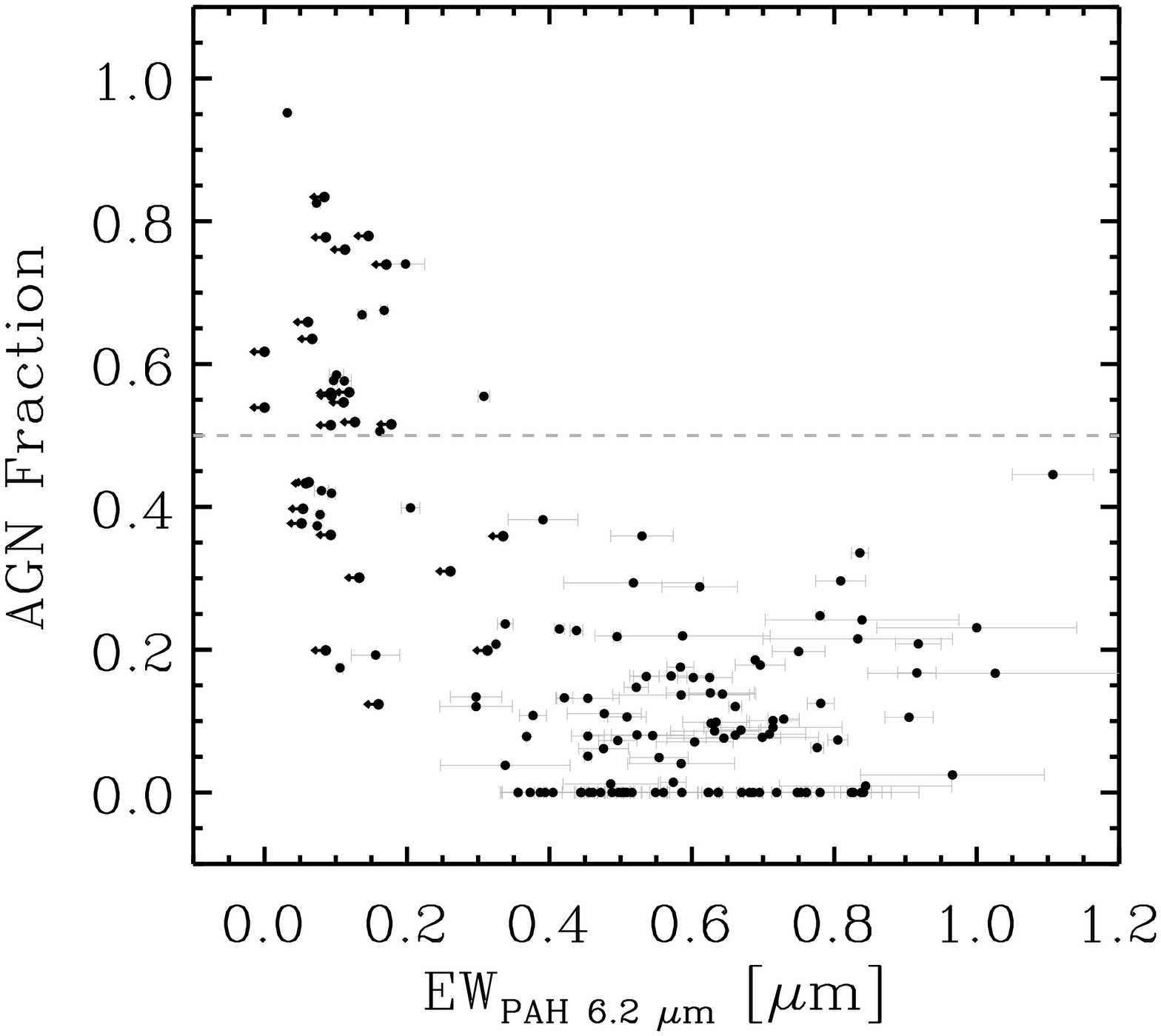}
\caption{{\textbf{\emph{Left}}}: IRAC colour-colour diagram of the 5MUSES sample. The wedge is the AGN selection region of Donley et al. (2012). Red and black circles  represent sources with $EW_{\rm 6.2} \leq 0.2\,\mu$m and $EW_{\rm 6.2} > 0.2\,\mu$m respectively. Grey circles correspond to sources for which the SED decomposition attributes more than 50\% of their total infrared luminosity to AGN activity.  
{\textbf{\emph{Right}}}: Fractional contribution of an AGN component to the total \lir\ as a function of the 6.2$\,\mu$m EW of the sources. The dashed grey line corresponds to $f_{\rm AGN} = 0.5$.}
\label{fig:fagn} %
\end{figure*}

\subsection{Classification of AGN and star-formation-dominated sources}
The equivalent widths of PAH features can serve as indicators of the AGN versus star formation activity in the galaxy (e.g. Laurent et al. 2000, Brandl et al. 2006, Spoon et al. 2007). Here, we used the 6.2$\,\mu$m equivalent width as measured by the spline-fit method. We chose the 6.2 over the 11.3$\,\mu$m PAH band because the latter is located on the shoulder of the 9.7$\,\mu$m silicate feature. Following previous studies (e.g., Armus et al. 2007; Wu et al. 2010), we classified sources with $EW_{\rm 6.2}$ $\leq$ 0.2$\,\mu$m as AGN-dominated and sources with $EW_{\rm 6.2}$ $> 0.2\,\mu$m as composite and star-forming- dominated sources, with purely star-forming galaxies having $EW_{\rm 6.2}$ $> 0.5\,\mu$m.

To validate the classification based on the 6.2$\,\mu$m PAH equivalent width we used the revised IRAC colour selection of Donley et al. (2012), which combines the AGN selection wedges (e.g., Lacy et al. 2004; Stern et al. 2005) with the  infrared power-law  selection of AGN (Alonso-Herrero et al. 2006; Donley et al. 2007; Park et al. 2010). This AGN selection limits the contamination by star-forming galaxies, and is also  reliable for the identification of luminous AGNs (Mendez et al. 2013). We found that all but one of the sources in our sample with $EW_{\rm 6.2}$ $>$ 0.2$\,\mu$m fail to meet the Donley et al. (2012) criteria, while $\sim$80\% of the sources with $EW_{\rm 6.2}$ $\leq$ 0.2$\,\mu$m would be classified as AGN based on their IRAC colours (Fig. \ref{fig:fagn} left).  

\begin{figure*}
\centering
\includegraphics[scale=0.7]{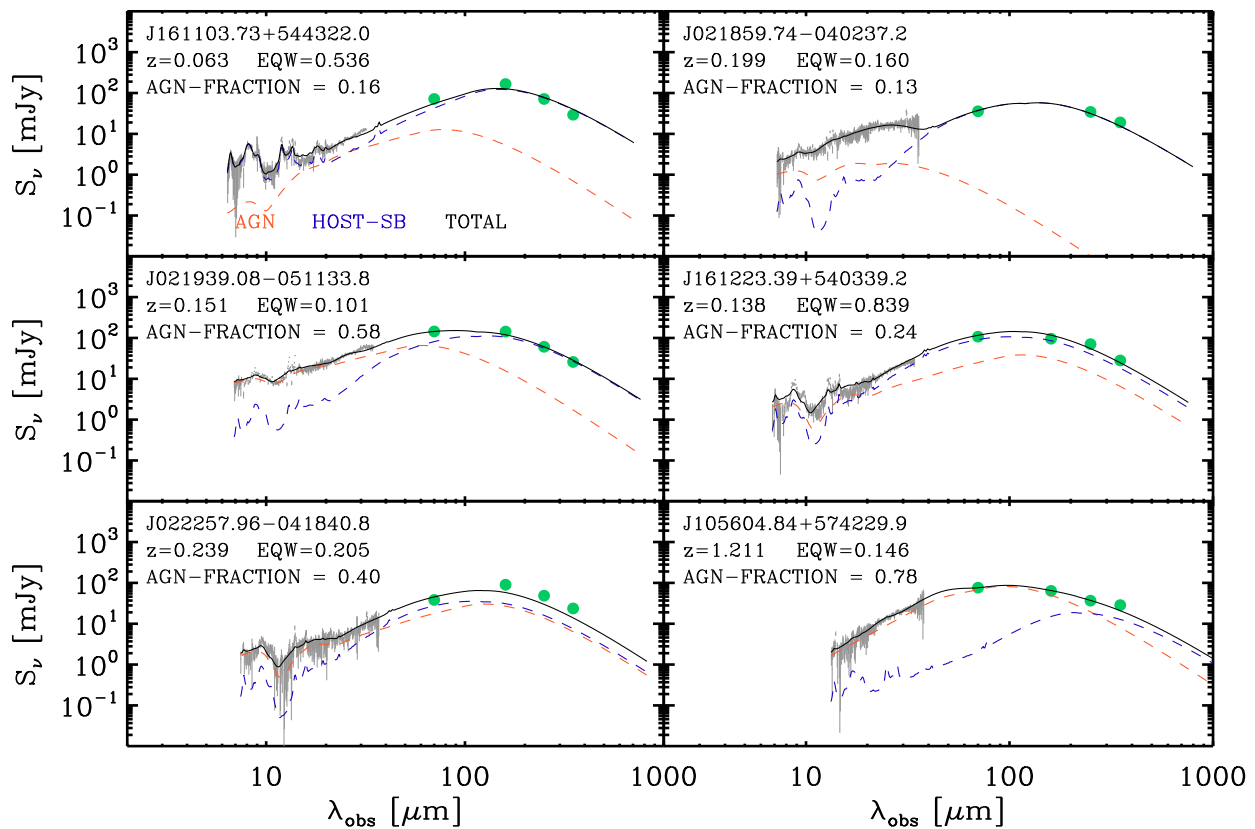}
\caption{Examples of SED fitting using the AGN/host-star-forming galaxy decomposition technique of Mullaney et al. (2011). Green circles are the observed points, overlaid with the best-fit total model (black line). The AGN and host galaxy components are shown as red and blue dashed lines, respectively.}
\label{fig:seds_agn} %
\end{figure*}

The plethora of mid-to-far-IR data that are available for our sample, and especially the IRS spectra, makes it possible to use a third  classification of the sources based on the  relative contribution of an AGN in the infrared output of our sample. For this task we use the AGN-host-galaxy decomposition method of 
Mullaney et al. (2011). This method employs a host-galaxy and an intrinsic AGN template SED to measure the contribution to the infrared output of  these two components{\footnote{We uses the DECOMPIR routine, available at \url{https://sites.google.com/site/decompir/}}}. This 
technique identifies the best-fitting model SED to the observed infrared data (spectra and photometry) through  $\chi^{2}$ minimisation and by 
varying the values of a set of free parameters. In brief, these free parameters are: 1) the host-galaxy SED (a set of five 
templates); 2) the wavelength of the  spectral break (if there is one) in the mid-IR SED of an AGN ($\lambda_{\rm brk}$); 3) the 
spectral indices ($\alpha_{\rm 1}$ and $\alpha_{\rm 2}$) below and above $\lambda_{\rm brk}$; 4) the wavelength at which the SED of the AGN 
component peaks; 5) the dust extinction of the AGN and host-galaxy component using a Draine et al. (2003) profile; and 6) the relative normalisations 
of the two components. Examples of the best model SEDs are shown in  Fig. \ref{fig:seds_agn}.

From the various output parameters here we focused on the recovered  \lir\ and the contribution of an AGN in the total infrared output ($f_{\rm AGN}$). The SED decomposition  can only trace possible misidentifications of AGN as star-forming galaxies, but not the other way round. Indeed, if our classification is correct, then the majority of the bolometric IR output of a star-formation-dominated galaxy cannot be due to AGN activity. On the other hand, a strong AGN that dominates the mid-IR spectrum  does not necessarily also dominate the total energy output of the source.  With this in mind, we plot in Fig. \ref{fig:fagn} (right)  the AGN fraction that corresponds to the best fit  as a function of $EW_{\rm 6.2}$. We found that sources with $EW_{\rm 6.2}$ $>$ 0.2$\,\mu$m tend to have, on average, a lower contribution of an AGN to their IR output, confirming the validity of our classification as star-forming dominated sources. In contrast, all sources for which we inferred that more than 50\% of the total \lir\ arises from dust heated by an AGN have $EW_{\rm 6.2}$ $\leq$ 0.2$\,\mu$m and the majority of them are found to meet the IRAC colour criteria of Donley et al. (2012) (Fig. \ref{fig:fagn} left). We also note that while for 60\% of the sources with $EW_{\rm 6.2}$ $>$ 0.2$\,\mu$m the best-fit yields a non-zero AGN contribution, based on  the Akaike information criterion (Akaike 1974),  a solution  without the need of an AGN component is equally probable within a 68\% confidence interval for 80\% of them. On the other hand, for almost all  sources (95\%) with $EW_{\rm 6.2}$ $<$ 0.2$\,\mu$m, the existence of an AGN is favoured at a $>$5$\sigma$ confidence level.

Given the very good agreement between the three independent indicators, we conclude that the EW of the 6.2$\,\mu$m feature is a reliable tool, at least for our sample, for identifying  AGN versus star-formation-dominated sources. Based on the  $EW_{\rm 6.2}$ classification, our sample consists of 116 SF and composite galaxies (70\%), and 50 AGN-dominated sources (30\%). While we did not find a trend between $f_{\rm AGN}$ and $EW_{\rm 6.2}$ in the $EW_{\rm 6.2} > 0.2\,\mu$m regime, for consistency with previous works in the literature we will refer to galaxies with 0.2$\,\mu$m $<$ $EW_{\rm 6.2}$ $\leq$ 0.5$\,\mu$m as composite and to galaxies with $EW_{\rm 6.2}$ $>$ 0.5$\,\mu$m as star-forming.
\section{Analysis}
Several key physical properties of distant galaxies, such as infrared luminosities (\lir), dust temperatures (\td) and dust masses (\md),  can be estimated by fitting their mid-to-far-IR SEDs with various models and templates. However, the lack of sufficient data for a proper characterisation of the SED has often limited this kind of analysis to models suffering from over-simplified assumptions and broad generalisations. The \s\ and \h\ data available for the galaxies in our sample provide thorough photometric sampling of their SEDs, allowing the use of more realistic models of the sort that have previously been applied mainly in the analysis of nearby galaxies. Here, we considered both the physically motivated Draine \& Li (2007, hereafter DL07) models for non-AGN-dominated sources and the more simplistic, but widely used, modified blackbody model (MBB) for the whole sample.
\begin{figure*}
\centering
\includegraphics[scale=0.75]{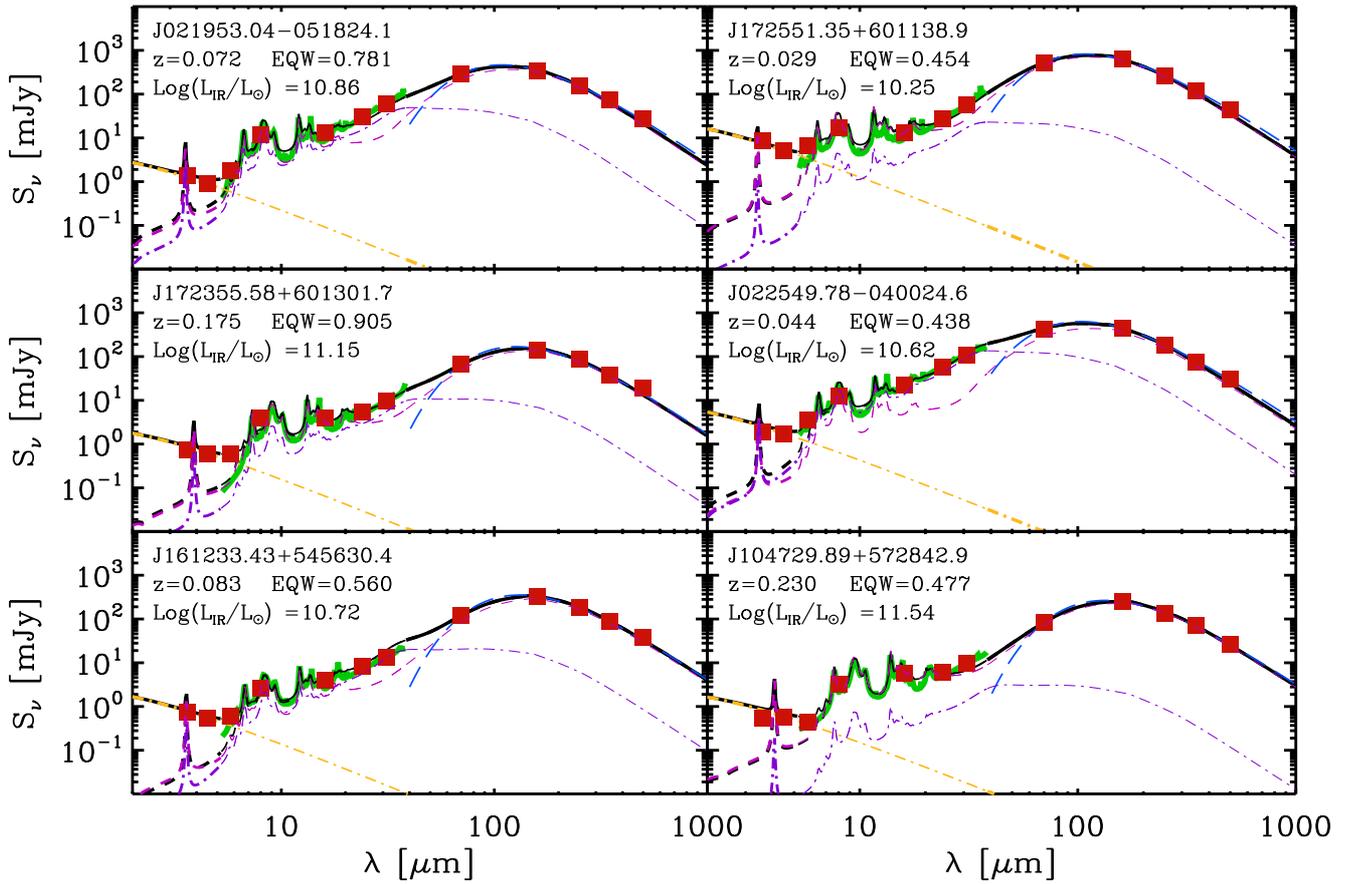}
\caption{Examples of SED fitting. Red squares are observed points, overlaid with the best-fit
DL07 model (black line). The PDR and diffuse ISM components are shown in purple. The yellow line is the stellar component and the dashed blue line the best-fit modified blackbody model with $\beta=1.5$. The green line in the mid-IR is the actual IRS spectrum of each source. We convolved the observed IRS spectrum with the IRS 16$\,\mu$m peak-up imaging filter and a 2$\,\mu$m-wide top-hat filter centred at  observed 30$\,\mu$m to enhance our fit with two more points in the observed mid-IR part of the spectrum. The rest of the data are taken from IRAC, MIPS, and SPIRE.}
\label{fig:seds} %
\end{figure*}
\subsection{Draine \& Li 2007 model}  
We  employed the dust models of DL07, which constitute an update of the models developed by Weingartner \& Draine (2001) and Li \& Draine (2001). These 
models describe the interstellar dust as a mixture of carbonaceous  and amorphous silicate grains, whose size distributions are chosen to mimic 
the observed extinction law in the Milky Way (MW), the Large Magellanic Cloud (LMC), and the Small Magellanic Cloud (SMC) bar region. The 
properties of these grains are parameterised by the PAH index, $q_{\rm PAH}$, defined as the fraction of the dust mass in the form of PAH grains. 
The majority of the dust is heated by a radiation field with a constant intensity $U_{\rm min}$, while a smaller fraction $\gamma$ of the dust is 
exposed to a power-law distribution of star light intensities extending from $U_{\rm min}$ to $U_{ \rm max}$. The $U = U_{\rm min}$ component can 
be  interpreted as the dust in the general diffuse ISM, while the power-law starlight distribution allows for dust heated by more intense 
starlight, such as in the intense PDRs in star-forming regions. For simplicity, emission from dust heated by $U >  
U_{\rm min}$ is referred to as the PDR, or the warm dust component, and the emission from dust heated by $U = U_{\rm min}$ is 
referred to as the diffuse ISM or the cold dust component. Although the PDR component contains only a small fraction of the total 
dust mass, in some galaxies it contributes a substantial fraction of the total power radiated by the dust. Then, according to DL07, the amount of 
dust, $d$\md, ~exposed to radiation intensities between $U$ and $U + dU$, can be expressed as a 
combination of a $\delta$-function and a power law:\\
\begin{equation}
\begin{centering}
{\rm \frac{dM_{\rm dust}}{dU} = \{(1-\gamma) M_{\rm dust} \delta(U-U_{\rm. min})+ \gamma M_{\rm dust} {\alpha-1 \over U_{\rm min}^{1-\alpha} - U_{\rm max}^{1-\alpha}} U^{-\alpha}},
\end{centering}
\end{equation}

\noindent with $U_{\rm min} \le U_{\rm max}, \alpha \ne 1$). Here, $U$ is normalised to the local Galactic interstellar radiation field, \md\ the total dust mass, $\alpha$ the power-law index, $\gamma$ the fraction  of the dust mass that is associated with the power-law part of the starlight intensity distribution, and  $U_{\rm min}$, $U_{\rm max}$ and $\alpha$  characterise the distribution of starlight intensities in the high-intensity regions. 

Following DL07, the spectrum of a galaxy can be described by a linear combination of one stellar component approximated by a  blackbody with colour temperature $T_*$ = 5000K, and two dust components, one arising from dust in the diffuse ISM, heated by a minimum radiation field $U_{min}$ (diffuse ISM component), and 
one from dust heated by a power-law distribution of starlight, associated with the intense  photodissociation regions (PDR component). Then, the model emission spectrum of a galaxy at distance $D$ is:

\begin{equation}
\begin{split}
f_{\nu}^{\rm model} = \Omega_* B_\nu(T_*) + {M_{\rm dust} \over 4 \pi D^2} \times & [ (1-\gamma) p_\nu^{(0)}(q_{\rm PAH},U_{\rm min}) + \\
&  \gamma p_\nu(q_{\rm PAH},U_{\rm min},U_{\rm max},\alpha)],\\
\end{split}
\end{equation}

\noindent where $\Omega_*$ is the solid angle subtended by stellar photospheres, $p_\nu^{(0)}(q_{\rm PAH}, U_{\rm min})$, and $p_\nu(q_{\rm PAH}$ $,U_{\rm min},U_{\rm max},\alpha)$ are the emitted power per unit frequency per unit dust mass for dust heated by a single starlight intensity $U_{\rm min}$, and dust heated by a power-law distribution of starlight intensities for $dM/dU \propto U^{-\alpha}$ extending from $U_{\rm min}$ to $U_{\rm max}$. 

In principle, the dust models in their most general form
are dictated by seven free parameters, ($\Omega_{*},  q_{\rm PAH}, U_{\rm min}, U_{\rm max}, \alpha, \gamma$ and $M_{\rm d}$). However, Draine et al.\ (2007) showed that the overall fit is insensitive to the details of the adopted dust model (MW, LMC, and SMC) and the precise values of $\alpha$ and  $U_{max}$. These authors showed that fixed values of  $\alpha=2$ and $U_{\rm max}=10^6$ successfully describe the SEDs of galaxies with 
a wide range of properties. Draine et al. 2007 also favour the choice of MW dust properties for which a set of models with $q_{\rm PAH}$ ranging from 0.4\% to 4.6\% is available. Furthermore, because low $U_{min}$ values correspond to dust temperatures below $\sim15$ K that cannot be constrained by far-IR 
photometry alone, in the absence of rest-frame (sub)mm data ($\lambda_{\rm rest}$ = 850$\,\mu$m), the authors using 0.7 $\le U_{\rm min} \le$ 25. 
While this lower cutoff for $U_{\rm min}$ prevents the fit from converging to erroneously large amounts of cold dust heated by weak starlight 
($U_{\rm min}$ $<$ 0.7), the caveat is a possible underestimate of the total dust mass if large amounts of cold dust are indeed present. 
However, Draine et al.\ (2007)  concluded that omitting rest-frame (sub)mm data from the fit increases the scatter of the derived masses to up to 
50\% but does not introduce a systematic bias in the derived total dust masses.

Under these assumptions, we fit the broadband \s\ and \h\ data of each galaxy with $EW_{\rm 6.2}$ $>$ 0.2$\,\mu$m in our sample, searching for the best-fit model by $
\chi^2$ minimisation and parametrising  the goodness of fit by the value of the  reduced $\chi^2$, $\chi^2_\nu \equiv \chi^2 / N_{\rm dof} $ (where $N_{\rm dof}$ is the number
 of degrees of freedom). To further exploit the available information  provided by the IRS spectra, for each source we estimated the observed flux density at 
 16$\,\mu$m as it would be measured with the 16$\,\mu$m IRS peak-up image, and at 30$\,\mu$m assuming a 2$\,\mu$m wide top-hat filter {\footnote{We 
 chose to measure the flux density at 30$\,\mu$m since beyond that wavelength the IRS spectra become progressively more noisy, as a result of the delimitation of the first-order long-low module filter.}}     

The best-fit model yields a total dust mass (\md), $U_{\rm min}$, as well as $\gamma$ and $q_{\rm PAH}$, while to derive \lir\footnote{$L_{\rm dust}$ quoted below is similar to $L_{\rm IR}$, but integrated from 0 to $\infty$} ~estimates we integrated the emerging SEDs from 8 to 1000$\mu$m:
\begin{equation}
  L_{\rm IR} = \int_{8~\mu m}^{1000~\mu m}  L_\nu (\lambda) \times \frac{c}{\lambda^{2}}~\rm d\lambda.
  \label{eq:LIR}
\end{equation}

A by-product of the best-fit model is also the 
dust-weighted mean starlight intensity scale factor, $\langle U \rangle$, defined as
\begin{equation}
  \langle U\rangle = \frac{L_{\rm dust}}{P_{0} M_{dust}}~ ,  \\
  \label{eq:avU}
\end{equation}
\noindent where $P_{0}$ is the power absorbed per unit dust mass in a radiation field with $U = 1$. Note that $\langle U \rangle$ is  essentially proportional to \lir/\md, and for the definition of \lir ~adopted here, that is, $L_{\rm 8-1000}$, Magdis et al. (2012b) have shown that $P_{0}$ $\approx$125. 

Uncertainties in \lir\ and \md\ were quantified using Monte Carlo simulations. 
To summarise, for each galaxy a Gaussian random number generator was used to create 1000 artificial flux sets from the original fluxes and measurement errors. These new data sets were then fitted in the same way, and the standard deviation in the new parameters was taken to represent the uncertainty in the parameters found from the real data set. Examples of the best-fit DL07 models along with the observed photometric points are shown in Fig. \ref{fig:seds},  while the set of best-fit parameters for each source are given in Table 2. We note that the DL07 models are representative of star-formation-dominated galaxies or equally for sources that do not harbour a strong AGN. As a consequence, our analysis was restricted to star-formation dominated galaxies. For AGN-dominated sources in our sample (i.e., $EW_{\rm 6.2}$ $\leq$ 0.2$\,\mu$m) we adopted the \lir\ measurements derived by the SED decomposition described in the previous section. 

\subsection{Comparison with modified blackbody fits}
Another method for deriving estimates of the dust properties is to fit the far-IR to submm SED of the galaxies with a single-temperature modified blackbody (MBB), expressed as
\begin{equation}
  f_{\rm \nu} \propto \frac{\nu^{3+\beta}}{e^{\frac{h\nu}{kT}}-1},
\end{equation}
\noindent where $T$ is the effective dust temperature (\td) and $\beta$ is the effective dust emissivity index.
Then, from the best-fit model, one can estimate  \md ~from the relation 
\begin{equation}
 M_{dust} = \frac{S_{\nu}D^{2}_{L}}{(1+z)\kappa_{\rm rest}B_{\nu}(\lambda_{\rm rest},T_{d})}, ~~with~~  \kappa_{\rm rest}=\kappa_{\rm 0}(\frac{\lambda_{o}}{\lambda_{\rm rest}})^{\beta}~,
\end{equation} 
\noindent where $S_{\rm \nu}$ is the observed flux density, $D_{\rm L}$ is the luminosity distance, and
 $\kappa_{rest}$ is the rest-frame dust mass absorption coefficient at the observed wavelength. 
While this is a simplistic approach, mainly adopted due to the lack of sufficient sampling of the SED
of distant galaxies, it has been one of the most widely used methods in the literature. Therefore, an analysis based on MBB-models provides both  estimates of the effective dust temperature of the galaxies in our sample, a quantity that is not directly measured from the DL07 model, and a valuable comparison between dust masses inferred with the MBB and DL07 methods. We note that unlike the DL07 analysis, which is restricted to non-AGN sources, the MBB technique and the derived \td\ measurements, are also valid for AGN-dominated sources.

We fit the standard form of a modified blackbody considering observed data points with $
\lambda_{\rm rest} > 60\,\mu$m, to avoid emission from very small grains, and used a fixed value of $\beta = 1.5$, typical of star-forming galaxies (Hildebrand 1983; Gordon et al. 2010; Magdis et al.\ 2011b; Dale et al. 2012). From the best fit model, we then estimated the total \md\ with equation 6, considering all sources, including AGN,  with at least three available photometric points at  
$\lambda_{\rm rest} > 60\,\mu$m.  For consistency with the DL07 models we adopted a value of $
\kappa_{\rm 250}$ = 5.1 cm$^{2}$ g$^{-1}$ (Li \& Draine 2001). To obtain the best-fit models and 
the corresponding  uncertainties of the parameters, we followed the same procedure as for the 
DL07 models. 
The derived parameters are summarised in Table 2 and the best-fit models are shown in Fig. \ref{fig:seds}

\begin{figure}
\centering
\includegraphics[scale=0.37]{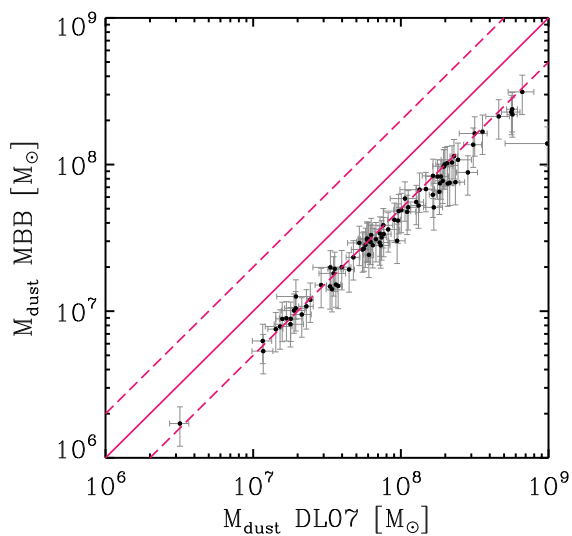}
\caption{Comparison between dust masses derived based on a single-temperature 
modified blackbody (MBB) and masses derived using the Draine \& Li (2007) models. For the MBB derived \md\ we assume $\beta = 1.5$. The purple solid line corresponds to unity  and the two purple dashed lines to its offset by a factor of 2 and 0.5.}
\label{fig:mdust} %
\end{figure}

A comparison between dust masses derived by DL07 and MBB is shown in Fig. \ref{fig:mdust}. We see that the  modified blackbody technique yields dust masses that are lower than those derived based on DL07 models on average by a factor of $\sim 2$ ($\langle M^{DL07}_{\rm dust}/M^{MBB}_{\rm dust}\rangle = 2.13 \pm 0.36$){\footnote{Fixing $\beta =$ 2, returns dust masses that are larger than those inferred when $\beta = 1.5$, by a factor of $\sim1.2$ but still lower by a factor of $\sim$ 1.8 than the masses derived by DL07}.  This result agrees with Magdis et al. (2012b) for a sample of z $\sim$ 1$-$2 star-forming galaxies and  Dale et al.\ (2012), who found that the discrepancy between the two dust mass estimates increases for sources with lower $S_{\rm 70}/S_{\rm 160}$ ratios that correspond to colder dust temperatures. This is caused by the inability of the single-temperature models to account for the wide range in the  temperature of dust grains  that are exposed to different intensities of  the interstellar radiation field. Fitting simultaneously both the Wien side of the modified black-body (which is dominated by warm dust), as well as the Rayleigh-Jeans tail (sensitive to colder dust emission) drives the derived temperatures to higher values and consequently to lower dust mass estimates (e.g. Dunne et al. 2000).  On the other hand, a more physically motivated  two-temperature blackbody fit returns dust masses that are larger by a factor of $\sim 2$ compared with those derived based on a single \td\ MBB (Dunne \& Eales 2001), in agreement with those inferred by the DL07 technique. For the rest of our analysis we only consider \md\ estimates derived based on the DL07 model.

\subsection{Comparison with \s\ \lir\ measurements}
In the pre-\h\ era the infrared properties of the 5MUSES sample were constrained based on \s\ IRAC, IRS, and MIPS data. In particular, for the redshift range of the majority of the sources, the 70$\,\mu$m and 160$\,\mu$m MIPS bands traced the emission of the galaxies on both sides of the peak of the SED, providing a first insight into the overall shape of their SEDs in the far-IR and their \lir\ values. However, only half of the sample is detected at 160$\,\mu$m. Furthermore, in the absence of the sub-mm data, the shape of the Rayleigh-Jeans tail of the galaxies in the 5MUSES sample was largely unconstrained. 

Here, we examined the impact of adding \h\ data to the derived \lir\ estimates of the sample.
In Fig. \ref{fig:lir} we compare the \lir\ measurements by Wu et al. (2010) based solely on  \s\ data (i.e., IRS and MIPS) with those derived in our study. The two estimates are in excellent agreement, with a mean ratio ($L^{\rm Herschel}_{\rm IR}/L^{\rm Spritzer}_{\rm IR}$) of 0.98 and a standard deviation of 0.22. This is in line with various studies that have shown  that  mid-IR extrapolations of the total \lir\ are correct for star-forming galaxies at  $z < 1.5$ (e.g., Elbaz et al. 2010; Rodighiero e al. 2010). We stress that only nine sources in our sample lie at $z > 1.5$, while $\sim$ 90\% of the 5MUSES sample considered here are at $z < 0.5$. The very good agreement between the two estimates also holds when we consider the three different classes of sources individually (starbursts, composite and AGN-dominated) although, we notice a somewhat larger discrepancy for the AGN-dominated sources, with  ($L^{\rm Herschel}_{\rm IR}/L^{\rm Spritzer}_{\rm IR}$) = 0.90 $\pm$ 0.35,  compared with 1.01 $\pm$ 0.16  and 0.99 $\pm$ 0.16 for composite and starburst galaxies, respectively. Finally, we note that the addition of \h\ submm data has a noticeable impact on the corresponding uncertainties of the measured \lir,  which are reduced by a factor of about 3.

\begin{figure}
\centering
\includegraphics[scale=0.4]{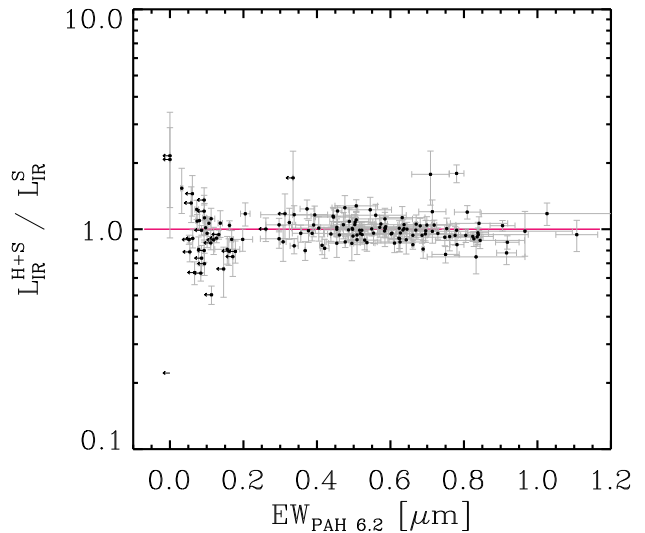}
\caption{Ratio of the total IR luminosity (\lir)  estimated using both \h\ and \s\ data ($L^{\rm H+S}_{\rm IR}$) over the one using only \s\ ($L^{\rm S}_{\rm IR}$, see Wu et al. 2010) as a function of the $EW_{\rm 6.2}$. (see Section 3.3). The solid  purple line corresponds to the 1$-$to$-$1 relation between the two \lir\ estimates.}
\label{fig:lir} %
\end{figure}

\section{Mid- to far-IR properties of AGN and star-forming galaxies}
The detailed mid-to-far-IR SEDs of our sample, as traced both by IRS spectroscopy and \s\ and \h\ broadband photometry, allows for an in-depth investigation of the total infrared spectra for our sample. In this section we  study possible correlations between mid-IR spectral features, and warm and cold dust components as well as variations of the PAH features between star-forming and AGN-dominated galaxies.   
\begin{figure*}
\centering
\includegraphics[scale=0.27]{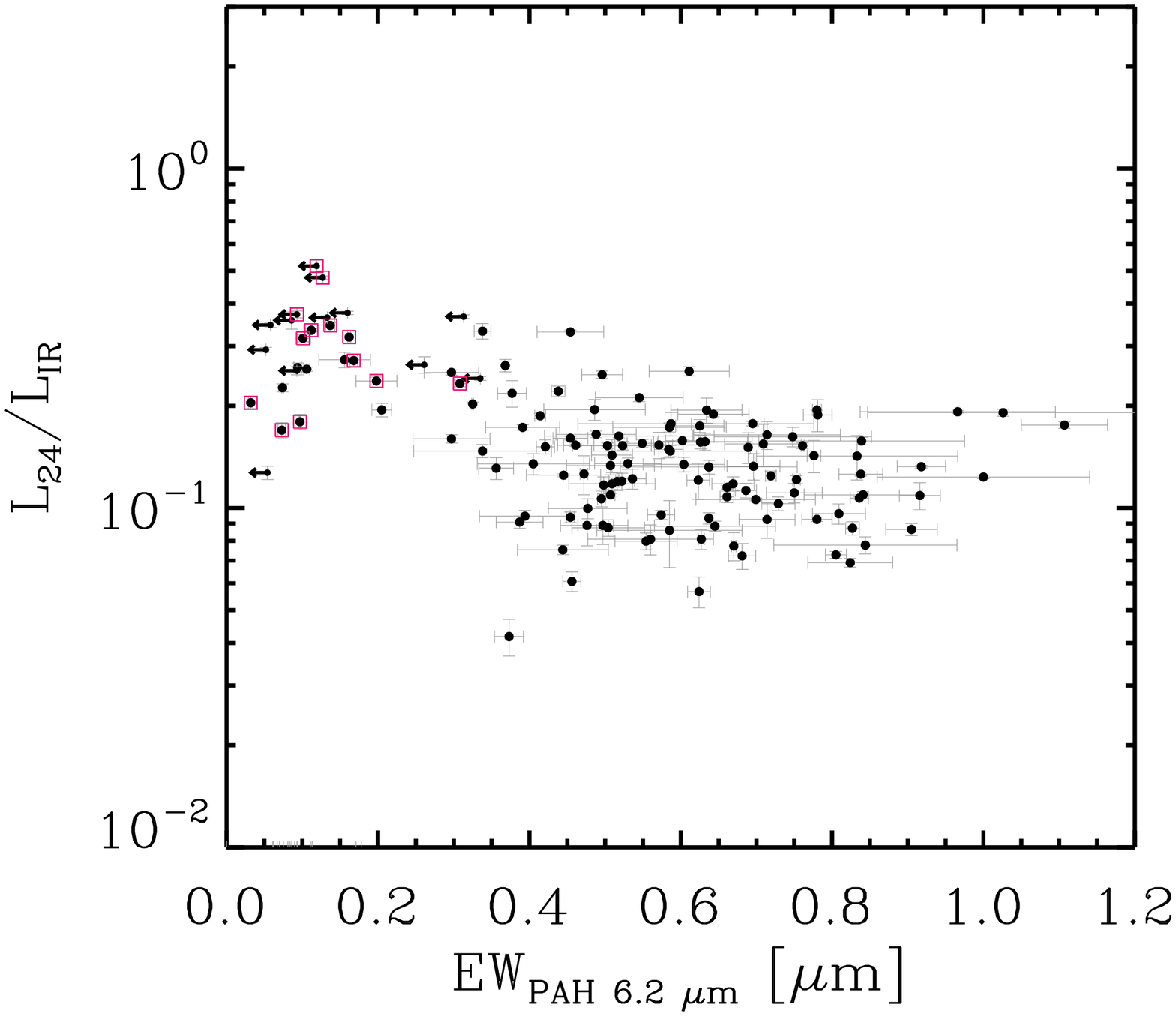}
\includegraphics[scale=0.27]{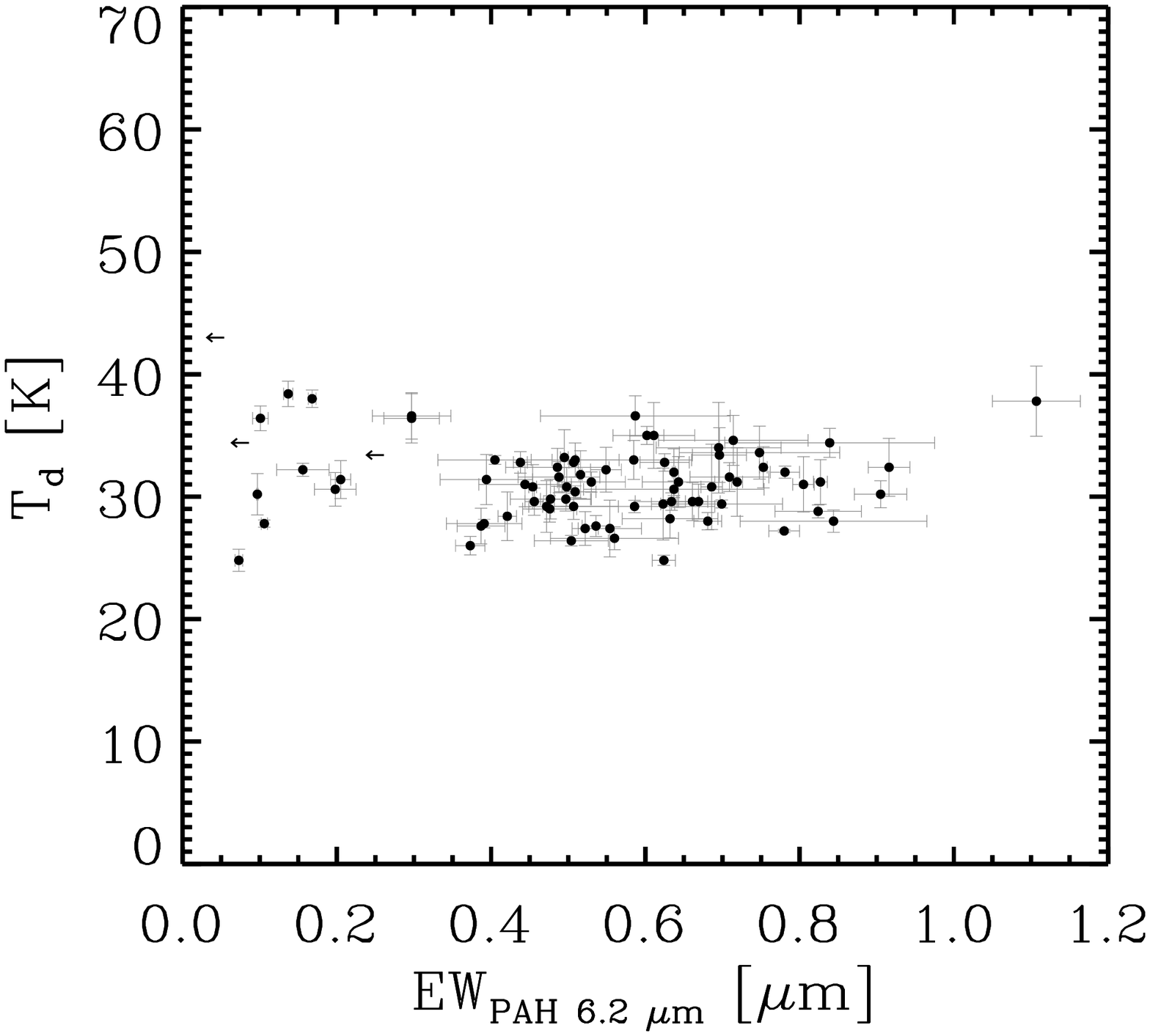}
\includegraphics[scale=0.27]{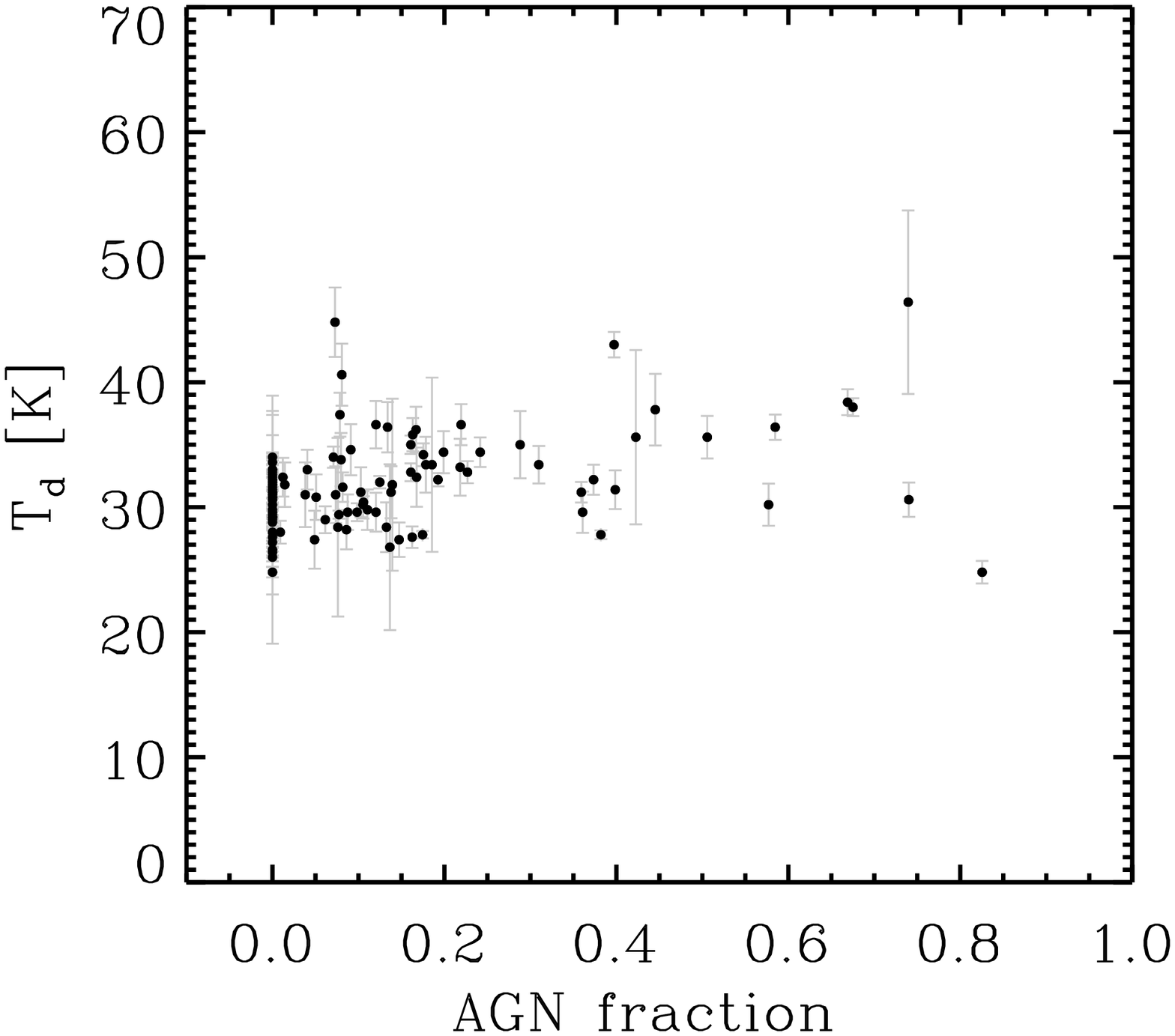}
\caption{{\textbf{\emph{Left}}}: $L_{\rm 24}$ / $L_{\rm IR}$ ratio as a function of the  6.2$\,\mu$m PAH feature EW. Purple squares correspond to sources for which the SED decomposition suggests that the AGN contribution to the total \lir\ is $> 50\%$. Leftward arrows denote upper limits in $EW_{\rm 6.2}$ measurements. {\textbf{\emph{Middle}}}: \td\ versus $EW_{\rm 6.2}$. \textbf{\td\ estimates are derived based on a modified blackbody model for 79 sources (out of the total 154 sources shown in the left panel), which are detected in at least three bands at $\lambda_{\rm rest}$ $>$ 60 $\,\mu$m, }.  {\textbf{\emph{Right}}}: \td\ versus fraction of \lir\ arising from AGN activity, as derived based on SED decomposition.}
\label{fig:td} %
\end{figure*}
\begin{figure*}
\centering
\includegraphics[scale=0.4]{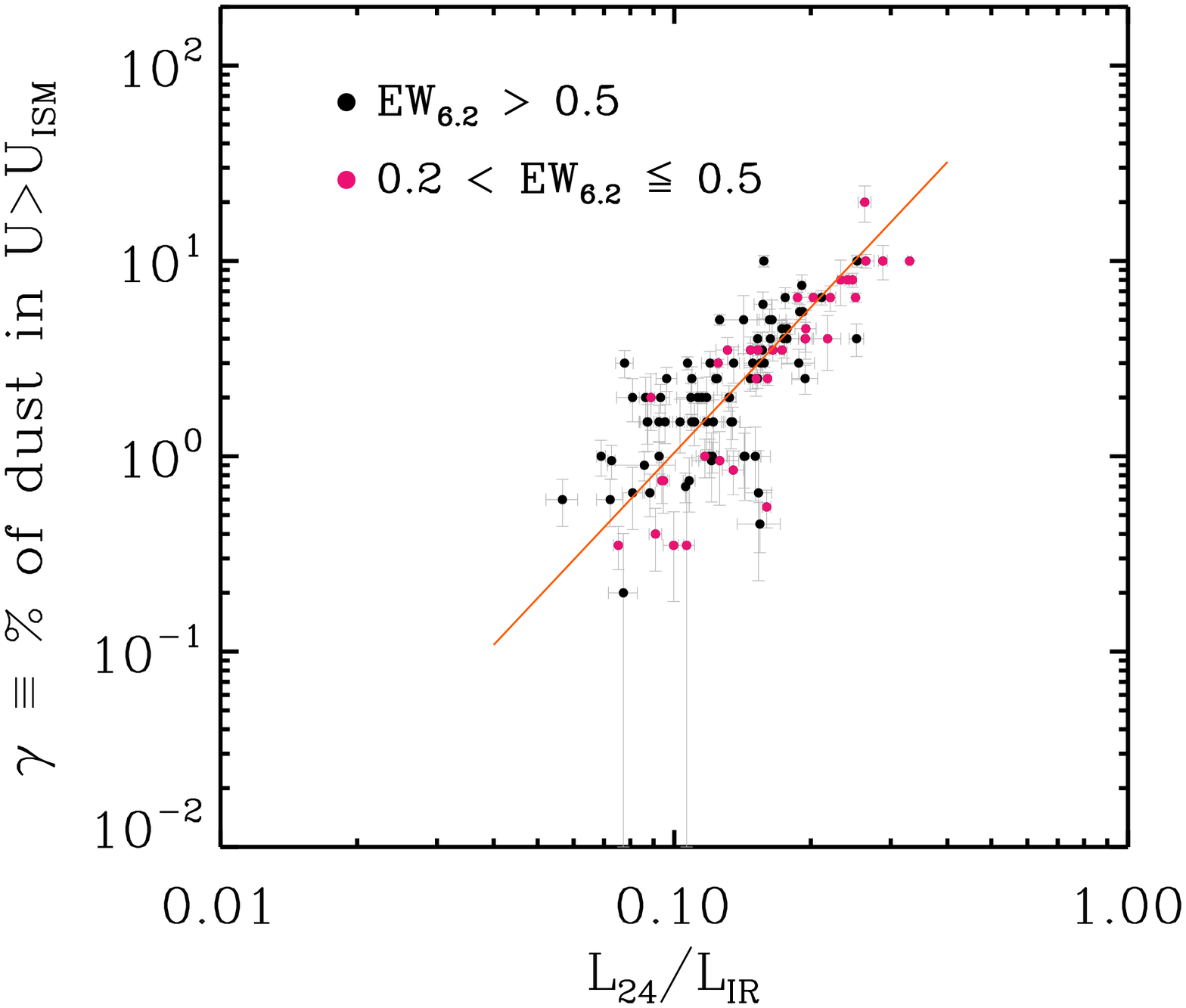}
\includegraphics[scale=0.4]{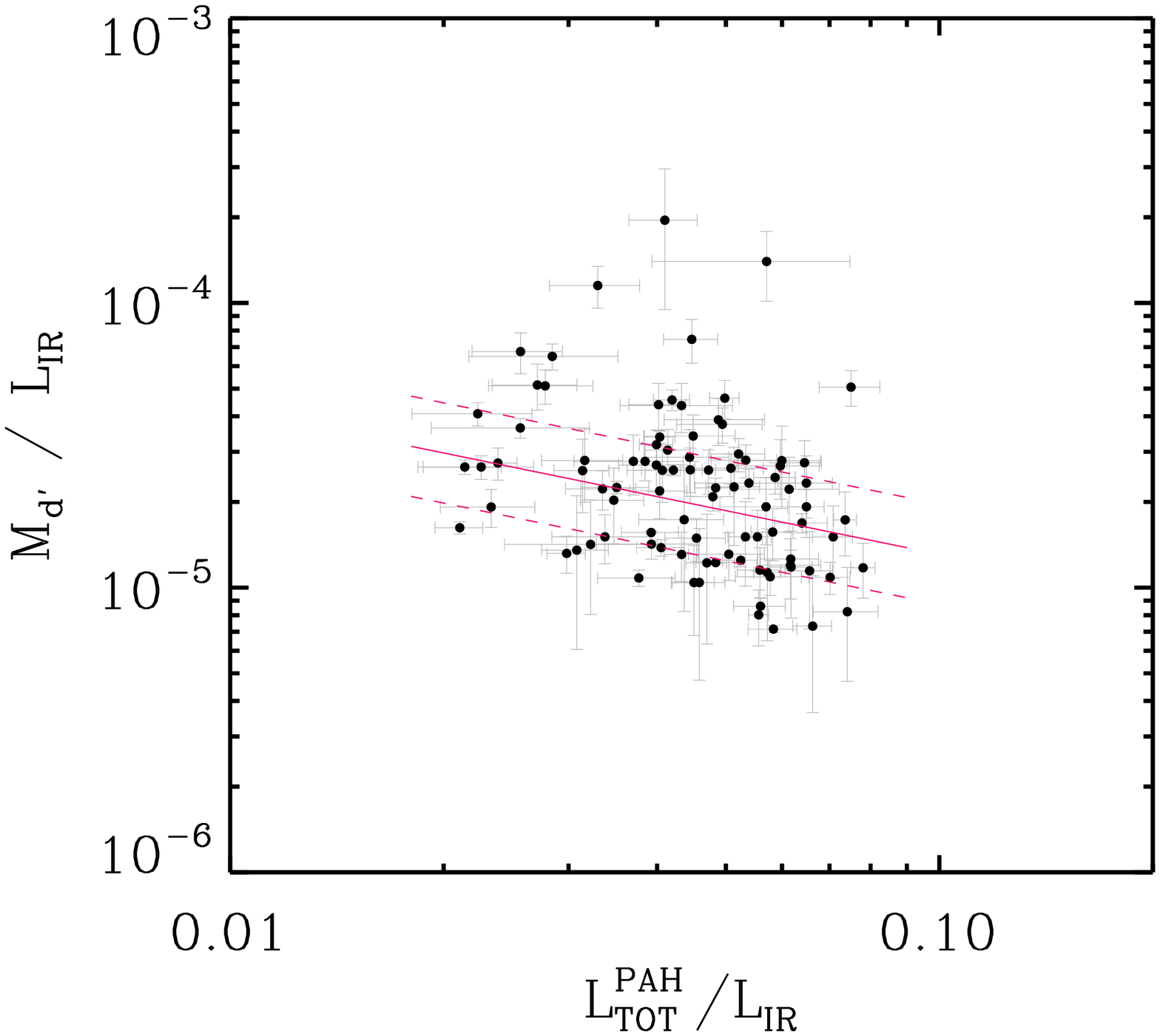}
\caption{{\textbf{\emph{Left}}}: Fraction of \md\ heated by radiation fields stronger than that of the diffuse ISM (or equivalently by $U > U_{\rm min}$) as a function of the $L_{\rm 24}$ / $L_{\rm IR}$ ratio for star-formation-dominated sources with $EW_{\rm 6.2}$ $>$ 0.5$\,\mu$m (black circles) and 
 0.2$\,\mu$m $<$ $EW_{\rm 6.2}$ $\leq$ 0.5$\,\mu$m (purple circles). The orange line depicts the best fit to the data. {\textbf{\emph{Right}}}: Dust mass heated by $U > U_{\rm min}$ versus $L^{\rm PAH}_{\rm TOT}$. Both quantities are normalised by \lir. The purple lines donate the best fit to the data and a scatter of 0.2$dex$.}

\label{fig:gamma} %
\end{figure*}
\subsection{Warm and cold dust}
The wide range in the 6.2$\,\mu$m EWs in our sample reveals a variety of mechanisms that power the warm dust emission, ranging from purely star-forming galaxies to AGN-dominated sources. The impact of  an AGN on the broadband mid-IR photometry is known to be prominent, with AGN-dominated sources exhibiting shallower mid-IR spectral slopes. For example, Wu et al. (2010) reported that while composite and SF sources share very similar $S_{\rm 30}/S_{\rm 15}$ (and $S_{\rm 70}/S_{\rm 24}$) flux density ratios, AGN-dominated sources are clearly separated from the rest with considerably lower $S_{\rm 30}/S_{\rm 15}$ (and $S_{\rm 70}/S_{\rm 24}$) values. With the addition of the \h\ data we are now in a position to advance this investigation by studying variations of the warm to cold dust emission among the different classes, and to infer the impact of an AGN on the total infrared  emission of the galaxies. 

We first investigated the $L_{\rm 24}$/$L_{\rm IR}$ ratio of our sources, where $L_{\rm 24}$ is the rest-frame 24$\,\mu$m luminosity as measured from the rest-frame IRS spectra of the galaxies. This ratio can serve as an indicator of the contribution of the warm emission to the total IR output of 
a galaxy, or equally to the relative amount of warm ($L_{\rm 24}$) to total ($L_{\rm IR}$) dust mass. As shown in Fig. \ref{fig:td} (left), AGN 
tend to exhibit an enhanced $L_{\rm 24}$/$L_{\rm IR}$ ratio compared with that of star-forming galaxies. We notice, however, that within each group (AGN, 
composites and star-forming galaxies) we find no correlation between $L_{\rm 24 }$/$L_{\rm IR}$ and $EW_{\rm 6.2}$. Instead, the $L_{\rm 24 }$/$L_{\rm IR}$  remains roughly constant within each group, albeit with a considerable scatter. In particular, we found a mean value and a standard deviation of $\langle L_{\rm 24}/L_{\rm IR} \rangle$ = 0.14 $\pm$ 0.04 for star-forming galaxies, 0.17 $\pm$ 0.08 for composites, and 0.32 $\pm$ 0.09 for AGN-dominated sources.
Evidently, the relative amount of warm dust in AGN as traced by the $L_{\rm 24 }$/$L_{\rm IR}$ ratio is higher than that found in star-forming 
galaxies, revealing an additional mechanism, to the energetic photons produced by young stars, which heats the dust and boosts the mid-IR 
emission. Because the relative amount of warm dust in AGN-dominated sources is boosted by a factor of $\sim$3 (for this redshift range), it is natural 
to expect that a considerable fraction of the bolometric infrared luminosity would arise from AGN activity in these systems. Indeed, based on the 
SED decomposition presented in the previous section, we found that for 75\% of the AGN with $L_{\rm 24}$/$L_{\rm IR}$ $>$ 0.3, more than half of the emerging \lir\ is powered by the AGN (see Fig. \ref{fig:td} left) . 

With the \h\ data in hand, we are also in a position to investigate whether the dominant powering 
mechanism, as traced by the spectral features in the mid-IR, has an impact on the cold dust
emission. In Fig. \ref{fig:td} (middle), we plot the derived \td\ measurements of the cold 
dust ($\lambda_{\rm rest} > 60\,\mu$m), as derived from the MBB fit, versus the 6.2$\,\mu$m EW of the sources in our sample. 
It appears that the cold dust temperature remains roughly constant for the whole range of EWs, 
with only a small, statistically insignificant, increase in \td\ as we move into the AGN regime. Indeed, composite and star-forming galaxies share similar dust temperatures with $\langle$ \td\ $\rangle$ $= (31 \pm 3)$ K, 
while AGN-dominated sources are marginally, although consistent within the uncertainties, warmer with $\langle$ \td\ $\rangle$ $= (34 \pm 4)$ K. 
The same trend is seen when we plot the dust temperature versus the fraction of \lir\ that 
arises from AGN activity within the galaxy (Fig. \ref{fig:td} right). Sources for which the 
SED decomposition suggests that more than half of the total \lir\ originates from AGN activity 
have $\langle$ \td\ $\rangle$ $= (34 \pm 3)$ K, while for the rest we found $\langle$ \td\ $\rangle$ 
$= (31 \pm 3)$ K. Hence, our analysis suggests that the far-IR part of the SED, as traced by 
the SPIRE bands in our sample, is predominantly shaped by, and directly linked to star 
formation and does not carry any measurable signature of AGN activity. This is in agreement with 
the findings of Hatziminaoglou et al. (2010), who reported that SPIRE colours, which 
are a good proxy of cold dust \td, are almost indistinguishable between star-forming and AGN-dominated sources. We stress though that this applies to AGN-dominated sources in our sample 
for which it was feasible to derive a  \td\ estimate, that is, those that are luminous enough in the submm to be detected  in the  SPIRE bands. With this in mind, we conclude that while the cold dust emission as traced by the SPIRE bands does not reveal the presence of an AGN, there is a marked difference in the mid- to far-IR colours of star-forming and AGN dominated sources, yielding larger amounts of warm dust in the latter. We note that similar results were reached by Kirkpatrick et al. (2012b) and Sajina et al. 2012, based on samples of $z = 0.4-5$ galaxies.

As we have seen, the bolometric infrared output of galaxies with $EW_{\rm 6.2} > 0.2\,\mu$m in our sample is dominated by 
star formation activity. Therefore, the derived parameters from the DL07 models provide meaningful constraints on the warm and cold dust emission. In Fig. \ref{fig:gamma}(left) we explore the contribution of a PDR component to the infrared output, 
or in other words the fraction of dust heated by PDRs ($\gamma$), as a function of the warm to total dust emission ratio  as 
traced by $L_{\rm 24}$/$L_{\rm IR}$. We recall that the ``PDR'' component in the DL07 models 
describes the amount of dust exposed to starlight with intensities higher than $U_{\rm min}$, which is the 
radiation field of the diffuse ISM. Both for the composite and for the star-forming galaxies, we found a  correlation 
between the two quantities, with the fraction of \lir\ that arises from PDRs increasing for galaxies with higher $L_{\rm 24}$/$L_{\rm IR}$ ratios, or equally for a larger portion of warm dust. Composite and star-forming galaxies follow the same trend and span the same $L_{\rm 24}/L_{\rm IR}$ range, pointing to a similar mechanism  heating their dust. 

In Fig.  \ref{fig:gamma}(right) we also investigate the variations of the relative contribution of the PAH features to the total \lir, as a function of the dust heated by radiation fields stronger than $U_{\rm min}$, that is,  $\gamma$ $\times$ \md. Instead of looking at a particular PAH emission feature we chose to consider the total PAH emission ($L^{\rm PAH}_{\rm TOT}$) as inferred by the sum of the fluxes of the 6.2, 7.7, and 11.3$\,\mu$m spectral features of our sources. We found for fixed \lir, a weak anti-correlation ($\rho = -0.38$) with PAH emission that decreases with increasing amount of dust heated by stronger radiation fields. While this trend could serve as evidence for PAH destruction in sources with a larger portion of their total dust mass exposed to stronger radiation fields, enhanced extinction or higher continuum levels cannot be ruled out by this analysis. In the next subsection we attempt to tackle this question.

 \begin{figure*}
\centering
\includegraphics[scale=0.4]{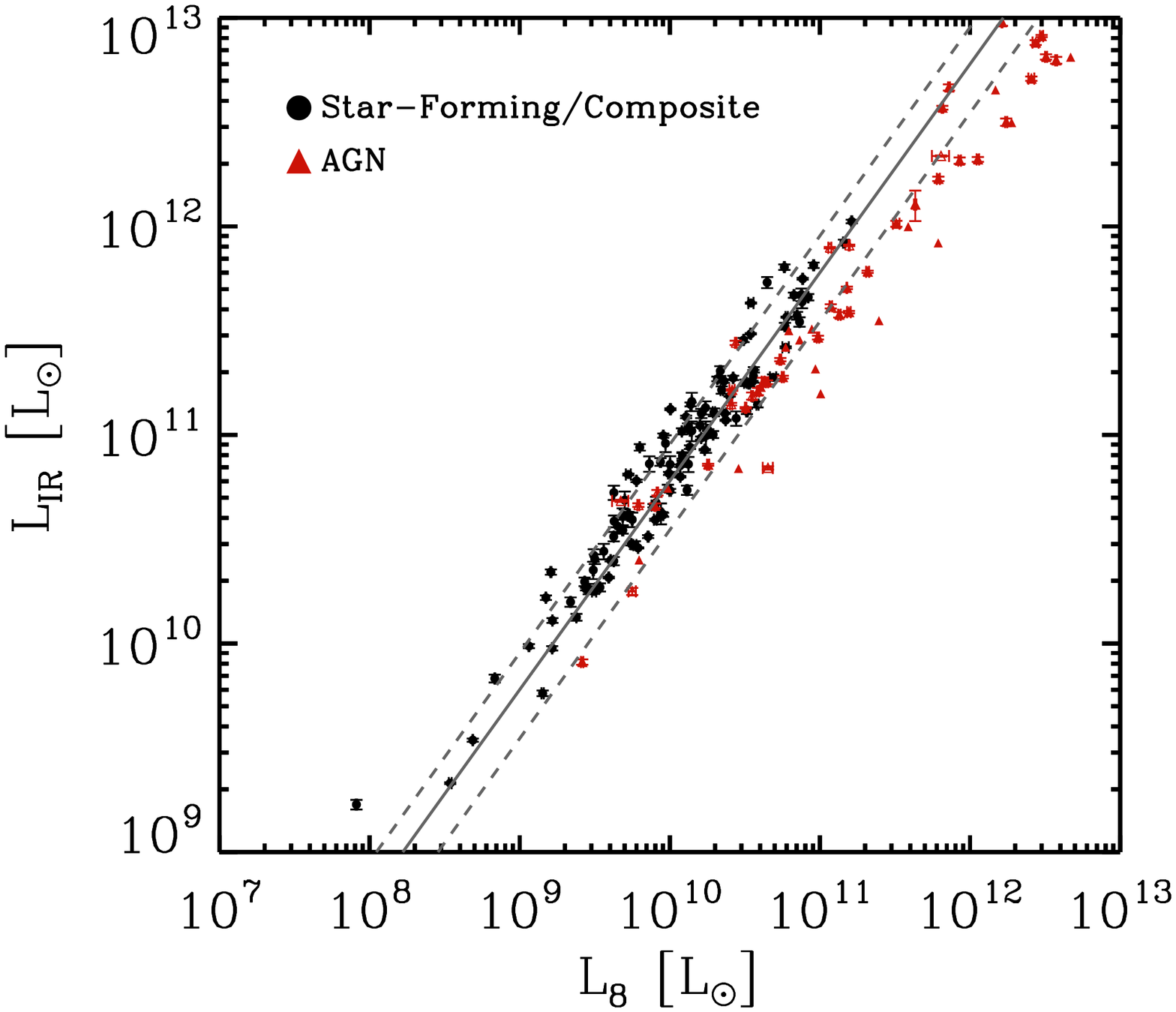}
\includegraphics[scale=0.4]{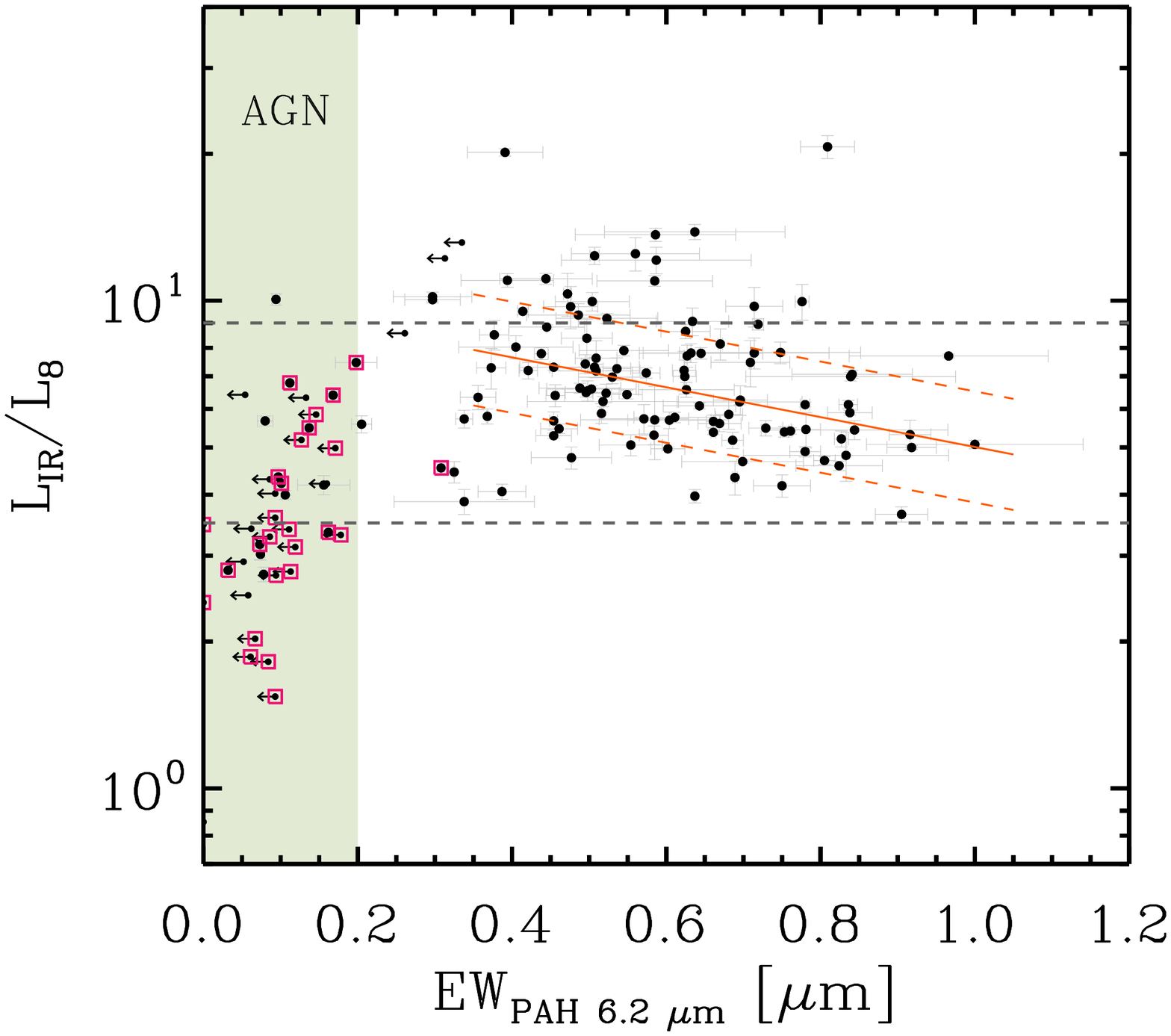}
\caption{{\textbf{\emph{Left}}}: \lir\ versus $L_{\rm 8}$ for the whole 5MUSES sample. Black circles correspond to purely star-forming and composite sources, while red triangles correspond to AGN. The solid grey line depicts the median and 68\% dispersion of composite and star-forming galaxies. {\textbf{\emph{Right}}}: \lir/$L_{\rm 8}$ versus EW of 6.2$\,\mu$m. The horizontal lines enclose the values of \lir/$L_{\rm 8}$ within the 68\% dispersion from the median. The orange lines depict the best-fit linear regression and its scatter in the $EW_{\rm 6.2} \geq$ 0.4$\,\mu$m regime. Purple squares correspond to sources with $f_{\rm AGN} > 0.5$.}
\label{fig:ir8} %
\end{figure*}

\subsection{\lir/\lmid\ versus PAH emission in star-forming galaxies} 
Recent studies based on \h\ data have revealed a scaling law for star-forming galaxies, relating the total IR luminosity, \lir, to the broadband (as traced by the IRAC 8$\,\mu$m filter) rest-frame 8$\,\mu$m luminosity, \lmid. In particular, Elbaz et al. (2011), showed that the 8$\,\mu$m bolometric correction factor, $IR8 \equiv$ \lir/\lmid, does not vary as a function of \lir\ or redshift. Instead, IR8 exhibits a Gaussian distribution containing the vast majority of star-forming galaxies both locally and up to $z \sim 2.5$, centred on IR8 $\approx$ 4.9 and with a scatter of $\sigma \approx 2.5$. However, since \lmid\ probes emission both from the 7.7$\,\mu$m complex and from the underlying continuum, it has been unclear whether the observed spread in the \lir$-$\lmid\ plane is mainly driven by PAH or mid-IR continuum variations. Furthermore, the determination of \lmid\ in these studies has primarily been based on K-corrected broadband photometry obtained by using template SEDs. Here, given the available IRS spectra, we can directly measure \lmid, and in combination with the accurate \lir\ estimates offered by the \h\ data, we can investigate the \lir$-$\lmid\ relation among star-forming galaxies as well as for sources ranging from purely star-forming to AGN-dominated.

To estimate \lmid\ for each source in our sample we convolved its rest-frame IRS spectrum with the IRAC 8$\,\mu$m filter.
To account for possible flux losses due to the narrow IRS short-low slit we applied a correction factor estimated from the the ratio of the photometric (broad-band) over the synthetic (IRS) observed 8$\,\mu$m (for sources with $z < 0.75$) and 24$\,\mu$m (for $z > 0.75$) flux density.  The overall correction is found to be on average $<$ 20\%. The derived  \lmid\ and \lir\ measurements are shown in Fig. \ref{fig:ir8} (left). For star-forming and composite galaxies ($EW_{\rm 6.2} >$ 0.2$\,\mu$m), the two luminosities are found to correlate almost linearly (slope of $\sim$ 0.97) with a median equal to IR8 $=$ 6.0 [-2.5,2.9] in the whole range of \lir\ (i.e. 10$^{9}$-$10^{12}$ \lsol). Interestingly, most of the AGN-dominated sources with \lir\ $< 10^{11.5}$ \lsol\ also lie within the 68\% scatter of the relation, in agreement with previous studies (e.g. Elbaz et al. 2011; Mullaney et al. 2012a; Kirkpatrick et al. 2012b). However, at higher luminosities (that also correspond to higher-redshift sources), the majority of AGN appear to deviate from the linear relation and  exhibit an enhanced \lmid\ for their \lir.

In Fig. \ref{fig:ir8}, we explore the variation of IR8 as a function of the 
6.2$\,\mu$m equivalent width. We identified two trends: a mild increase in the IR8 as we move 
from purely star-forming to composite sources, and a dramatic drop when we enter 
the AGN regime ($EW_{\rm 6.2}$ $< 0.2\,\mu$m). In the $EW_{\rm 6.2}$ $\geq$ 0.4$\,\mu$m regime a Spearman test yields a 
statistically significant ($p = 2.8 \times 10^{-5}$) anti-correlation ($\rho = -0.41$) between IR8 
and $EW_{\rm 6.2}$. To examine whether this correlation is robust against the uncertainties associated with the $EW_{\rm 6.2}$ and IR8 values, 
we created 1000 realisations by allowing the data to take values selected from Gaussians (given the measured values and errors) and repeated the correlation ranking test. The mean and  
standard deviation of the derived correlations are $\rho = -0.39 \pm 0.09$, suggesting a weak but statistically significant 
anti-correlation between the two parameters. 

Interestingly, this trend is not driven by \lir\ because in this regime ($EW_{\rm 6.2} > 0.2\,\mu$m), we found that IR8 and $EW_{\rm 6.2}$ do not correlate with \lir. Since the $EW_{\rm 6.2}$ is an indicator of the strength of the 6.2$\,\mu$m line relative to the underlying continuum, we examined whether this  behaviour could be driven by variations of the  6.2$\,\mu$m PAH features or by variations of the mid-IR continuum level. For a fixed \lir, \lmid\ would increase if the drop in the $EW_{\rm 6.2}$ had been caused due to an elevated continuum, and hence IR8 would remain constant or even drop. Therefore, the decrease in $EW_{\rm 6.2}$ for star-forming galaxies is  more likely to be due to lower PAH emission relative to \lir, rather than an elevated 8$\,\mu$m continuum emission. On the other hand, the sharp decrease of IR8 at $EW_{\rm 6.2}$ $< 0.2\,\mu$m, can be understood as an increase in the mid-IR continuum that results in both higher \lmid\ and lower $EW_{\rm 6.2}$ for a given \lir.  
\begin{figure*}
\centering
\includegraphics[scale=0.4]{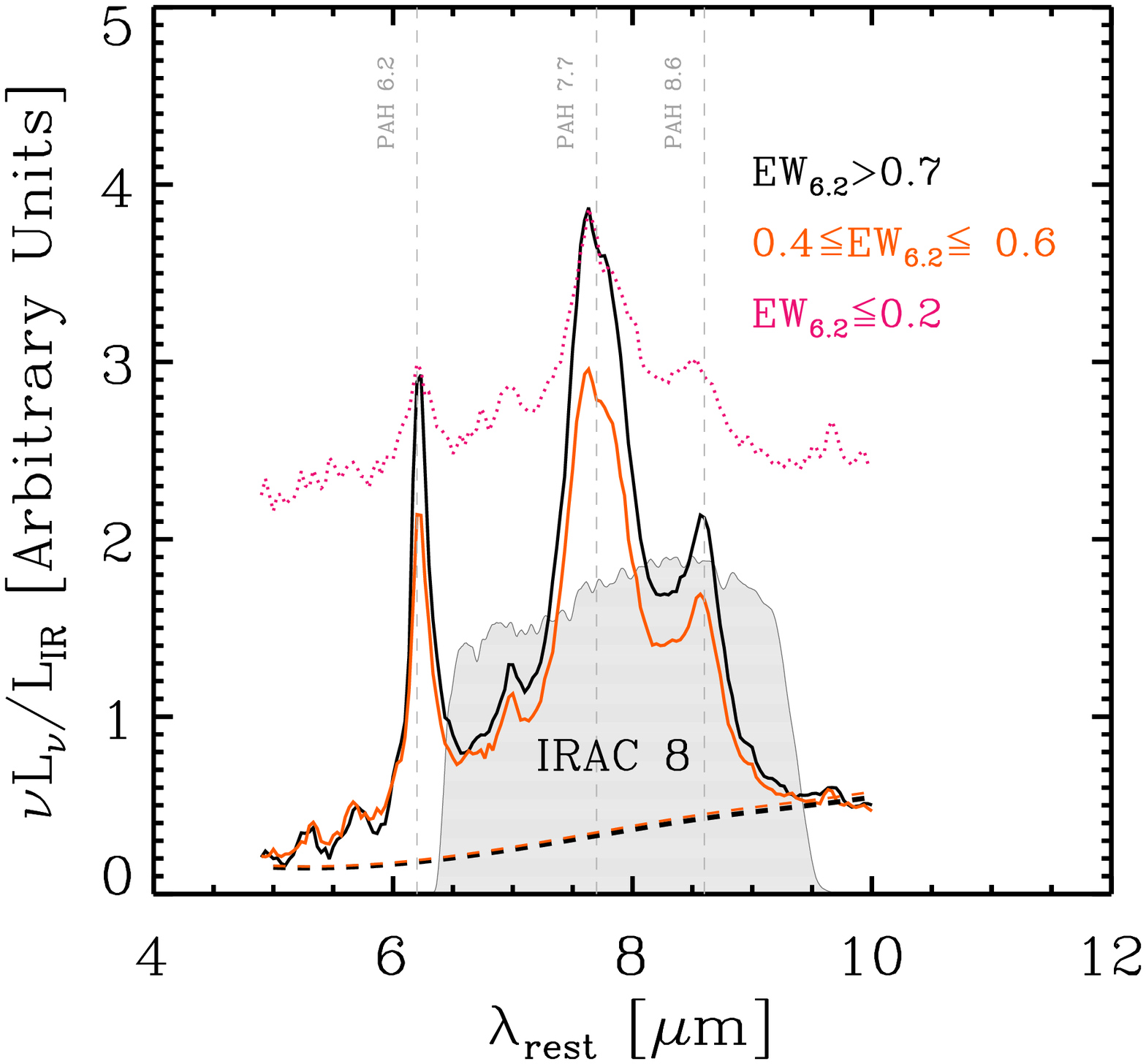}
\includegraphics[scale=0.4]{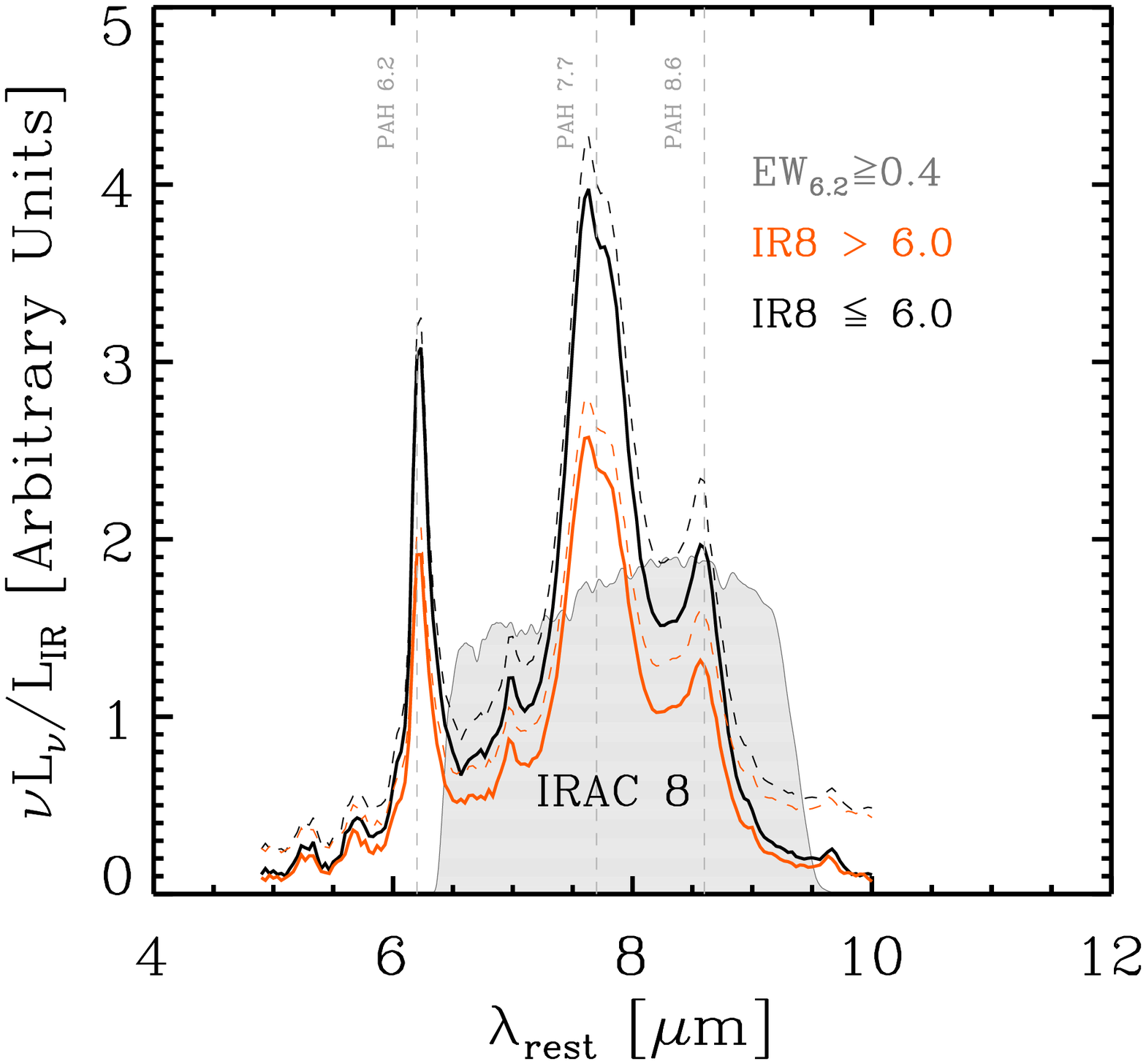}
\caption{{\textbf{\emph{Left}}}: Average spectra in various $EW_{\rm 6.2}$ bins in the $4-10\,\mu$m range. The dashed lines correspond to the underlying continuum emission as inferred  by PAHFIT. The grey shadowed area depicts the 8.0$\,\mu$m IRAC filter. {\textbf{\emph{Right}}}: Average (dashed lines), and average continuum subtracted (solid lines) spectra of of all galaxies with $EW_{\rm 6.2}$ $\geq$ 0.4$\,\mu$m grouped in those with IR8 $>$ 6 (orange lines), and IR8 $<$ 6  (black lines)}
\label{fig:spec} %
\end{figure*}

To explore this scenario we split the sources into two bins of $EW_{\rm 6.2}$, $0.4\,\mu$m $\leq$ $EW_{\rm 6.2}$ $\leq$ 0.6$\,\mu$m and $EW_{\rm 6.2}$ $>$ 0.7$\,\mu$m. We then constructed a median IRS spectrum for each bin by stacking the individual spectra normalised to \lir\ $=$ 1 \lsol. To ensure that the rest-frame 5$\,\mu$m emission was traced by the IRS spectra we placed a lower limit of $z = 0.07$ on the redshift of the sources that
 enter the stack. For each  spectrum we also define a local continuum or plateau below the PAH features by fitting the PAH features with 
 Drude profiles using PAHFIT as described in Smith et al. (2007). The resulting spectra in the rest-frame 5$-$11$\,\mu$m  wavelength range and 
 the corresponding continuum levels are shown in Fig. \ref{fig:spec} (left). The two spectra exhibit almost identical continua at 6.2$\,\mu$m 
 with a noticeable difference in the emission of the  PAH feature. This implies that for fixed \lir, a decrease in the $EW_{\rm 6.2}$ of star$-
 $forming galaxies is primarily caused by a decrease in the emission of the 6.2$\,\mu$m PAH feature and not by dilution from an elevated 
 continuum emission in the 5$-$15$\,\mu$m regime. On the other hand, the mean IRS spectrum of AGN dominated sources ($EW_{\rm 6.2}$ $\leq$ 0.2$\,\mu$m) 
 normalised to \lir$ = 1$ \lsol\ exhibits a mid-IR continuum about 10 times higher than that of the star-forming galaxies, suggesting that the 
 decrease in IR8  is due to PAH destruction or smearing from the strong continuum emission. We note that repeating the analysis using a spline 
 function does not change our result. 
 
The mean spectra in various $EW_{\rm 6.2}$ bins also provide clues about the apparent increase of IR8 with decreasing $EW_{\rm 6.2}$ among star-forming galaxies. As shown in Fig. \ref{fig:spec} (left), \lmid\ probes both the 7.7$\,\mu$m PAH complex and the continuum emission arising from hot dust. Since the presented spectra are normalised to the same \lir\, we can directly infer that at fixed \lir, sources with higher $EW_{\rm 6.2}$ have higher \lmid\ and hence, lower IR8. Indeed, measuring \lmid\ directly from the spectra yields an IR8 value for the sources in the 0.4$\,\mu$m $\leq$ $EW_{\rm 6.2}$ $\leq$ 0.6$\,\mu$m bin higher by a factor of $\sim$1.4 than to those with $EW_{\rm 6.2}$ $\geq$ 0.7$\,\mu$m. The variation of the PAH features as a function of IR8 is more directly seen when we split our sample of star-forming galaxies into two IR8 bins, above and below IR8 $=$ 6.0. The continuum-subtracted average IRS spectra of the two sub-samples are presented in Fig. \ref{fig:spec} (right), clearly demonstrating a decrease of the PAHs emission for sources in the high IR8 bin. Indeed, the emission from the PAH features measured directly from the stacked spectra is higher by a factor of $\sim$1.5 in the low IR8 bin than that in the high IR8 bin. 

We also investigated whether this result could be an artefact because different galactic scales are probed by the IRS at different redshifts. The low-z cutoff that we introduced in the stack ensures that the smallest galaxy scale traced by IRS is $\sim$6 kpc. Therefore, because we only probe a fraction of the disk for the low$-$z sample, we could indeed be missing extended PAH emission (e.g. Pereira-Santaella et al. 2010; Diaz-Santos et al. 2010b).
An indication for the maximum of the PAH emission that we might be missing is given by the ratio of the broad-band (which traces the whole galaxy) 
to spectral flux ratio. As discussed above, this is only at a $\sim$15\% level and therefore, our result would still hold because the observed PAH variation is of a factor of $\sim$1.5. We note though that this extreme scenario is unlikely to be the case because we would also see a correlation between the 6.2$\,\mu$m EW (or IR8) and redshift. The absence of this correlation (as indicated from our data), suggests that the missed flux 
cannot originate solely from PAHs. Furthermore,  Elbaz et al. (2011)  and Diaz-Santos (2011) have shown that sources with higher IR8 exhibit 
more extended 11.3$\,\mu$m PAH emission. Accordingly,  one would expect that PAH emission is more likely to be missed for sources with lower IR8 values, which would additionally  strengthen our result. Finally, repeating the analysis without applying any correction to \lmid\ yields the same 
results, albeit with an increase of the mean IR8 from 6.0 to 6.2. In summary, we have provided evidence that the spread in IR8 values of star-forming galaxies mirrors variations in their PAH emission. In the next section we discuss possible explanations for the origin of this observation.

\section{Discussion}
We have seen that the dispersion in the \lir$-$\lmid\ relation defined by star-forming galaxies traces variations in the PAHs emission with respect to \lir. Because the  \lir/\lmid\ and  \lir/$L_{\rm PAH}$  do not appear to correlate with \lir\ (see also Wu et al. 2010), other  physical parameters must drive this variation. 

A possible explanation could be a varying level of AGN activity. However, there are several arguments against this scenario. First, for almost all sources with $EW_{\rm 6.2}$ $\geq$ 0.4$\,\mu$m the SED decomposition allocates the bulk of \lir\ to star-formation activity and none of them meet the power-law AGN criterion of Donley et al. (2012), known to be a reliable selection of AGN (Mendez et al. 2013). A prominent AGN activity would also boost \lmid\ and therefore  decrease IR8 for a given \lir. This becomes  clear in the $EW_{\rm 6.2} <$ 0.2$\,\mu$m regime, where almost all sources are selected as power-law AGNs and, as stated above, lie in the lower regime of the IR8 main sequence or fall below it. Finally, as shown in Fig.  \ref{fig:fagn}(right), we do not observe any correlation between $EW_{\rm 6.2}$ and contribution of an AGN to the total \lir. This means that the observed increase of IR8 with decreasing $EW_{\rm 6.2}$ cannot be explained by an increasing AGN activity. 
 \begin{figure*}
\centering
\includegraphics[scale=0.39]{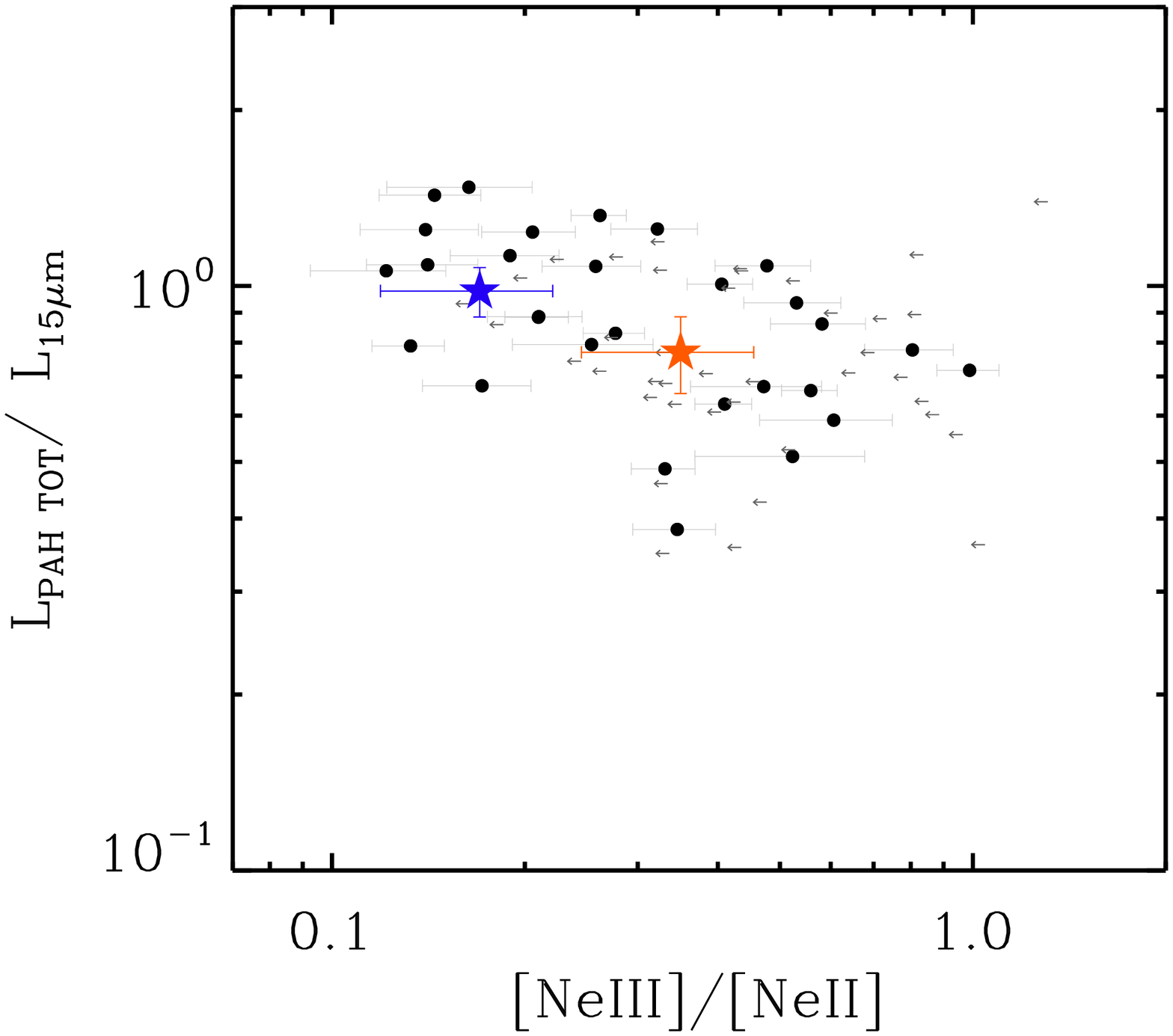}
\includegraphics[scale=0.39]{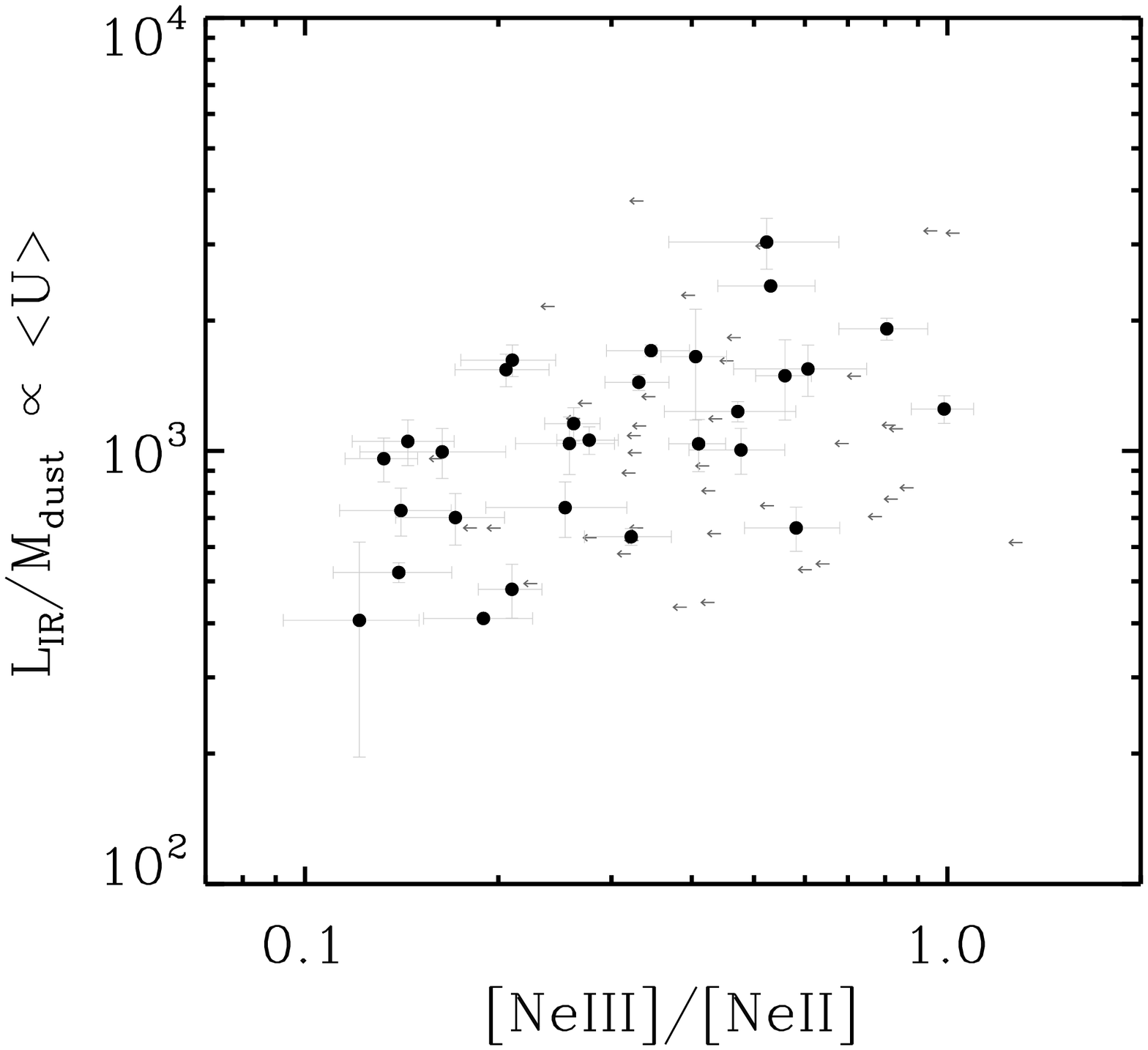}
\caption{{\textbf{\emph{Left}}}: $L^{\rm PAH}_{\rm TOT}$/$L_{\rm 15}$ versus [$\ion{Ne}{III}$]/[$\ion{Ne}{II}$], as measured from the IRS spectra of the sources in our sample with $EW_{\rm 6.2} \geq$ 0.4$\,\mu$m. The orange and blue stars correspond to the average spectra of galaxies with IR8 $\leq$ 6.0 and 6.0 $<$ IR8 $<$ 9.0, respectively. Leftward arrows denote upper limits of the [$\ion{Ne}{III}$]/[$\ion{Ne}{II}$] ratio for sources with only a [\ion{Ne}{II}] detection {\textbf{\emph{Right}}}: Dust-mass-weighted luminosity (\lir/\md) as a function of the hardness of the radiation field traced by the  [$\ion{Ne}{III}$]/[$\ion{Ne}{II}$] line ratio.}
\label{fig:ne} %
\end{figure*}

 An alternative scenario that could explain the variation of the PAH strength among star-forming galaxies is a variation in the hardness of the interstellar radiation field ($U$). As argued by Abel et al. (2009), under the assumption of a constant gas density, an increase in $U$ and therefore of the number of  ionising photons ($h\nu > 13.6$eV) produced by young, massive, OB stars, would lead to an increase of the ratio of ionised to atomic hydrogen and consequently to a reduced  gas opacity. As $U$ increases, the \ion{H}{II} regions extend to higher A$_{V}$ into the cloud, and a larger portion of the UV photons are absorbed by the dust in the ionised region and are re-emitted in the form of IR emission. The net effect is that the fraction of UV photons available to ionise or excite the PAH molecules in the surrounding PAH-rich PDRs or in the neutral ISM is reduced for higher $U$ values, resulting in lower PAH emission for a fixed \lir\ (e.g. Peeters et al 2004). In this scenario one would expect an anti-correlation between the strength of the PAHs relative to the continuum emission and the hardness of the radiation field. This has already been shown to be the case in various samples of local galaxies and star-forming regions in the local Universe (e.g. Peeters et al. 2002, Madden et al. 2006, Wu et al. 2006). 
 
 To test this scenario  we considered the ratio of [$\ion{Ne}{III}$] $\lambda$15.56$\,\mu$m and the [$\ion{Ne}{III}$] $\lambda$12.81$\,\mu$m lines, which is a common tracer of the hardness of the radiation field (e.g. Genzel et al. 1998; Madden et al. 2006; Farrah et al. 2007). The diagnostic value of this ratio is based on 1) the large difference in the ionisation potential of the Ne$^{++}$(41 eV) and Ne$^{+}$(21.6 eV), 2) the fact that this ratio is independent of the neon abundances, and 3) does not suffer from differential extinction due to the proximity of the wavelength  of the two emitted lines. In Fig. \ref{fig:ne} (left) we plot the $L^{\rm PAH}_{\rm TOT}/L_{\rm 15}$ as a function of  [$\ion{Ne}{III}$]/[$\ion{Ne}{II}$] for galaxies in our sample with $EW_{\rm 6.2}$ $\geq$ 0.4$\,\mu$m and for which the two lines have been detected. The anti-correlation between the strength of the radiation field and the relative strength of the PAH seen in previous studies of local galaxies appears to hold for our sample, too, suggesting that galaxies with harder radiation fields, as traced by the [$\ion{Ne}{III}$]/[$\ion{Ne}{II}$] line ratio, tend to exhibit weaker PAH emission than the underlying continuum. In this scenario the PAH deficit could originate from PAH destruction from energetic photons produced by young and massive stars. 
 
 We also measured the [$\ion{Ne}{III}$]/[$\ion{Ne}{II}$] ratio from the stacked spectra in the two IR8 bins (IR8 $<$ 6.0 and 6.0$<$ IR8 $<$ 9.0) and found that galaxies with higher IR8 values have a mean [$\ion{Ne}{III}$]/[$\ion{Ne}{II}$]  (and $L^{\rm PAH}_{\rm TOT}/L_{\rm 15}$) ratio that is approximately a factor of $\sim$2 higher than that of galaxies with IR8 $<$ 6.0. We note, however, that there is only a very weak trend between IR8 and [$\ion{Ne}{III}$]/[$\ion{Ne}{II}$]  in individual detections. To explain this lack of a clear correlation between IR8 and [$\ion{Ne}{III}$]/[$\ion{Ne}{II}$] we recall that  [$\ion{Ne}{III}$]/[$\ion{Ne}{II}$] might mirror variations in the age of the starburst (e.g. Thornley et al. 2000). It is possible that galaxies with higher [$\ion{Ne}{III}$]/[$\ion{Ne}{II}$] ratios have experienced a more recent (up to 2Myrs) starburst event with a larger fraction of young and massive  OB stars than older starbursts. Therefore, the mid-IR fine structures  lines (those of neon in our case) predominantly trace the youngest stellar populations ($\sim$ few Myr). On the other hand, the total infrared luminosity (the one of the two parameters that shape IR8) traces emission by dust heated by stellar population up to  $\sim 100$~Myr.  As a consequence, changes in the line ratios can occur on much shorter timescales than changes in IR8, if the starburst has been occurring for more than 10 Myr. That could also offer an alternative explanation for the variation of IR8 as a function of $EW_{\rm 6.2}$; IR8 variations could result from a mechanism that affects \lir, but not the PAH emission, such as a contribution of an old stellar population to \lir. That would suggest that galaxies with higher IR8 values are galaxies that have experienced an older starburst event and their bolometric infrared output, \lir, significant contributes by dust heated by old stellar populations.  However, as shown by Elbaz et al. (2011), high IR8 values in the local Universe, but also at high-$z$,  are predominantly found among galaxies that are experiencing a recent star-burst event with high specific star formation rates (such as  local ULIRGs and SMGs). 
 \begin{figure*}
\centering
\includegraphics[scale=0.6]{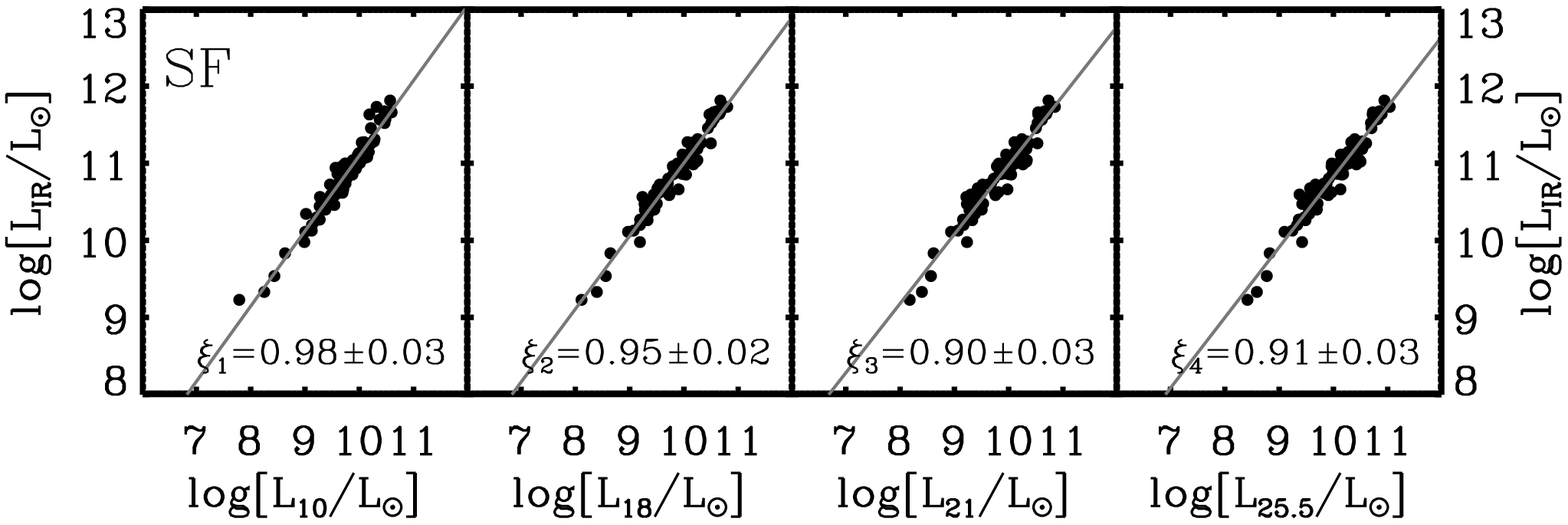}\\
\includegraphics[scale=0.6]{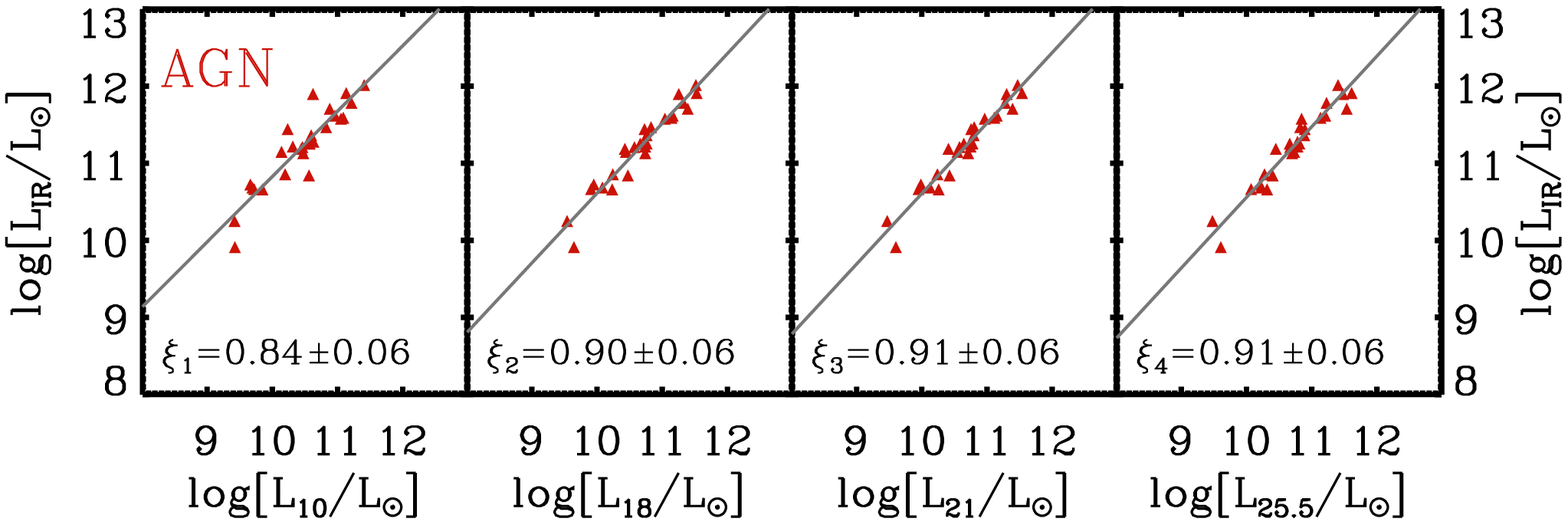}
\caption{Correlation between \lir\ and rest-frame luminosities at observed 10, 18, 21-, and 25.5$\,\mu$m.
The choice of the rest-frame luminosity wavelengths corresponds to the central wavelengths of the JWST-MIRI filters.
Each IRS spectrum was shifted to the rest-frame and  convolved with MIRI filters. The best-fit regression line and the associated plot are shown in each panel and are summarised in Table 3.}
\label{fig:jwst} %
\end{figure*}
In the above framework, a harder radiation field would also result in a larger number of UV photons per dust particle, or equally in a higher dust-mass-weighted luminosity, \lir/\md. Indeed, this quantity reflects the amount of available energy per dust mass unit, under the assumption that the majority of the UV photons are eventually absorbed by dust and are re-emitted in the IR. We note that in the DL07 models the dust-mass-weighted luminosity is proportional to the mean radiation field ($\langle U \rangle$) and is a good proxy of the dust temperature of the ISM of the galaxy (Magdis et al. 2012b). As shown in Fig. \ref{fig:ne} (right), \lir/\md\ correlates with [$\ion{Ne}{III}$]/[$\ion{Ne}{II}$], albeit with a considerable scatter, suggesting that the dust-mass-weighted luminosity (or \td) traces  the hardness of the radiation field in star-forming galaxies.  

Differential line extinction due to gas density variations can also  cause a decrease in the observed [$\ion{Ne}{III}$]/[$\ion{Ne}{II}$] ratio. For example, Farrah et al. (2007)  argued for an increased density of gas in local ULIRGs based on their lower [$\ion{Ne}{III}$]/[$\ion{Ne}{II}$] ratios (for a given [$\ion{S}{IV}$]/[$\ion{S}{III}$])  compared with those of  systems with lower infrared luminosities. However, a sole increase in gas density cannot explain  the observed \lir/\md$-$[$\ion{Ne}{III}$]/[$\ion{Ne}{II}$] correlation in our sample. Finally, metallicity effects could also play a role. Galaxies with lower IR8 or PAH emission could simply be more metal-rich and therefore have more PAHs, resulting in higher \lmid\ for a given \lir. While [$\ion{Ne}{III}$]/[$\ion{Ne}{II}$] is not expected to be directly affected by metallicity variations, various studies have shown that  the [$\ion{Ne}{III}$]/[$\ion{Ne}{II}$] ratios in low-metallicity galaxies are very high ($>$3), often one to two orders of magnitude greater than in more metal-rich starburst galaxies (e.g., Rigby \& Rieke 2004, Madden et al. 2006,  Wu et al. 2006). Metal poor galaxies also tend to exhibit lower (by a factor of  10) $L^{\rm PAH}_{\rm TOT}/L_{\rm 15}$ ratios than  metal-rich starbursts (Madden et al. 2006). While not conclusive, the fact that our sample exhibits only  a narrow range in the [$\ion{Ne}{III}$]/[$\ion{Ne}{II}$] (0.1$<$[$\ion{Ne}{III}$]/[$\ion{Ne}{II}$]$<$0.8) and $L^{\rm TOT}_{\rm PAH}$ values,  and does not extend to values similar to those of local metal poor dwarfs suggests that metallicity is not the main driver of the observed dispersion.

While a firm conclusion cannot be reached, it appears that the PAH deficit with respect to \lir\ in star-forming galaxies with higher IR8 values is caused by PAH destruction due to harder radiation fields that also result in higher dust temperatures. 
These characteristics could be indicative of a more compact star-formation geometry, which is in line with the recently proposed correlation between IR8, compactness of the projected star-formation density, and excess in the specific star formation rate (e.g., Diaz Santos et al. 2011; Elbaz et al. 2011). However, a proper study of the spatial extent of the PAH emission and of the star-forming regions in distant galaxies as traced by the mid-IR emission will have to wait for the launch of the James Webb Space Telescope (JWST). Given that a large number of distant galaxies that will be detected with MIRI onboard JWST will lack detection in longer wavelengths it will be extremely useful to have scaling relations between the mid-IR and total infrared luminosity. Using the observed IRS spectra of the galaxies and the total infrared luminosity as inferred from the addition of the \h\ data, we provide  total luminosity calibrations based on monochromatic luminosities at 10-, 18-, 21- and 25.5-$\,\mu$m as they will be traced by MIRI for AGN-dominated and star-forming galaxies (Fig. \ref{fig:jwst} and Table 3). 

\section{Conclusions}
We have used a 24$\,\mu$m-selected flux-limited sample of 154 $z \sim 0.15$ galaxies all chosen to benefit from ancillary IRS spectra and new \h\ SPIRE photometry as part of the HerMES programme to construct their full IR SEDs. Based on SED decomposition, on the relative strength of the 6.2$\,\mu$m PAH feature with respect to the mid-IR continuum, and on the IRAC colours of the sources, we have identified the dominant mechanism that powers their infrared emission, classified them as AGN and star-forming-dominated galaxies and investigated in detail their mid-to-far-IR properties. Our main findings are summarised as follows:
\begin{itemize}
  
\item We found a statistically insignificant effect of the presence of an AGN on the temperature of the cold dust of the host galaxy. AGN-dominated sources exhibit a marginally higher cold dust temperature, suggesting that the far-IR colours are mainly shaped by star formation activity.
\item Star-forming galaxies show an anti-correlation between the IR8=\lir/\lmid and the 6.2$\,\mu$m PAH feature $EW_{\rm 6.2}$. Our analysis suggests that for galaxies with $EW_{\rm 6.2}$ $\geq$ 0.4$\,\mu$m (i.e. star-forming galaxies), differences in the $EW_{\rm 6.2}$, for fixed \lir, are driven by variations in the PAH emission and not by a varying 5$-$15$\,\mu$m underlying continuum.  
\item For star-formation-dominated sources we confirm that the strength of the PAH emission anti-correlates with the hardness of the radiation field as traced by the [$\ion{Ne}{III}$]/[$\ion{Ne}{II}$] ratio. 
\item The dust-mass-weighted luminosity \lir~/ \md, correlates with the [$\ion{Ne}{III}$]/[$\ion{Ne}{II}$] ratio, suggesting that  sources with harder interstellar radiation fields are characterised by higher dust temperatures.
\end{itemize} 

Overall, it appears that the decrease in the PAH strength, for fixed \lir, is caused by harder radiation fields that are characterised by high dust-mass-weighted luminosities and dust temperatures. While this is in line with recent evidence that link the IR8 with the compactness of the star-forming regions, the upcoming JWST mission will allow for a more detailed investigation both in local and in distant galaxies.

\textit{Acknowledgments.} GEM acknowledges support from the John Fell Oxford University Press (OUP) Research Fund and the University of Oxford. A.A.-H. acknowledges funding through the Universidad de Cantabria
August G. Linares Programme. VC acknowledges partial support from the EU FP7 Grant PIRSES-GA-20120316788. This paper uses data from \h's photometer SPIRE. SPIRE has been developed by a consortium of institutes led by Cardiff University (UK) and including Univ. Lethbridge (Canada); NAOC (China); CEA, LAM (France); IFSI, Univ. Padua (Italy); IAC (Spain); Stockholm Observatory (Sweden); Imperial College Lon- don, RAL, UCL-MSSL, UKATC, Univ. Sussex (UK); and Caltech, JPL, NHSC, Univ. Colorado (USA).

\bibliographystyle{aa} % style aa.bst
\bibliography{ms_v3}

\onecolumn
\begin{deluxetable}{lccccc}

\tabletypesize{\normalsize}
\tablewidth{0pc}
\tablecaption{\h\ SPIRE Photometry of the 5MUSES sample}
\tablehead{

\colhead{Name} &
\colhead{$z_{\rm spec}$} &
\colhead{$S_{\rm 24}$} &
\colhead{$S_{\rm 250}$} &
\colhead{$S_{\rm 350}$} &
\colhead{$S_{\rm 500}$} \\

\colhead{} &
\colhead{} &
\colhead{[mJy]} &
\colhead{[mJy]}&
\colhead{[mJy]}&
\colhead{[mJy]}} 
\startdata
 J021503.52-042421.6  &   0.137  &    5.2$\pm$0.3&   19.4$\pm$2.4  &   -  &  -  \\
 J021557.11-033729.0  &   0.032  &    8.8$\pm$0.4&  123.5$\pm$2.4  &   58.2$\pm$2.6  &   21.0$\pm$3.4  \\
 J021649.71-042554.8  &   0.143  &   10.1$\pm$0.5&   33.5$\pm$2.5  &  -  &  -  \\
 J021743.82-051751.7  &   0.031  &   17.1$\pm$0.9&  190.9$\pm$2.4  &   89.9$\pm$2.6  &   44.9$\pm$3.0  \\
 J021754.88-035826.4  &   0.226  &   10.3$\pm$0.5&  138.2$\pm$2.3  &   83.2$\pm$2.5  &   41.1$\pm$3.1  \\
 J021849.76-052158.2  &   0.292  &    5.3$\pm$0.3&   61.7$\pm$2.4  &   20.7$\pm$2.5  &  -  \\
 J021859.74-040237.2  &   0.199  &   15.9$\pm$0.8&   34.5$\pm$2.4  &   19.0$\pm$2.4  &  -  \\
 J021916.05-055726.9  &   0.103  &   11.0$\pm$0.6&   45.7$\pm$2.8  &   25.3$\pm$3.0  &   20.2$\pm$3.4  \\
 J021928.33-042239.8  &   0.042  &   17.3$\pm$0.9&   43.9$\pm$2.5  &   21.7$\pm$2.4  &  -  \\
 J021939.08-051133.8  &   0.151  &   32.5$\pm$1.6&   60.6$\pm$2.1  &   25.8$\pm$2.2  &  -  \\
 J021953.04-051824.1  &   0.072  &   30.3$\pm$1.5&  157.6$\pm$2.2  &   72.9$\pm$2.2  &   27.8$\pm$2.8  \\
 J021956.96-052440.4  &   0.081  &    5.6$\pm$0.3&   70.8$\pm$2.5  &   38.0$\pm$2.3  &  -  \\
 J022005.93-031545.7  &   1.560  &    6.9$\pm$0.3&   37.5$\pm$2.4  &   38.5$\pm$4.7  &  -  \\
 J022145.09-053207.4  &   0.008  &    6.2$\pm$0.3&   36.5$\pm$2.1  &   22.0$\pm$2.2  &  -  \\
 J022147.82-025730.7  &   0.068  &   21.0$\pm$1.1&  218.2$\pm$2.4  &   96.0$\pm$2.9  &   27.0$\pm$4.4  \\
 J022147.87-044613.5  &   0.025  &    5.1$\pm$0.3&   20.3$\pm$2.5  &  -  &  -  \\
 J022151.54-032911.8  &   0.164  &    6.9$\pm$0.3&   51.1$\pm$2.5  &   27.6$\pm$2.8  &  -  \\
 J022205.03-050537.0  &   0.258  &    6.3$\pm$0.3&   96.5$\pm$2.4  &   47.1$\pm$2.7  &   22.5$\pm$4.7  \\
 J022223.26-044319.8  &   0.073  &    5.1$\pm$0.3&   40.3$\pm$2.4  &  -  &  -  \\
 J022224.06-050550.3  &   0.149  &    5.7$\pm$0.3&   33.5$\pm$2.5  &  -  &   30.8$\pm$2.8  \\
 J022241.34-045652.0  &   0.139  &    5.1$\pm$0.3&   30.6$\pm$2.4  &  -  &  -  \\
 J022257.96-041840.8  &   0.239  &    5.3$\pm$0.3&   48.5$\pm$2.3  &   23.6$\pm$2.5  &  -  \\
 J022301.97-052335.8  &   0.708  &    6.8$\pm$0.3&  141.8$\pm$2.4  &   73.3$\pm$2.4  &   39.1$\pm$3.1  \\
 J022315.58-040606.0  &   0.199  &    9.4$\pm$0.5&   76.3$\pm$2.1  &   38.4$\pm$2.5  &   24.4$\pm$2.6  \\
 J022329.13-043209.5  &   0.144  &    7.6$\pm$0.4&   50.0$\pm$2.5  &  -  &  -  \\
 J022334.65-035229.4  &   0.176  &    7.6$\pm$0.4&   25.8$\pm$2.3  &  -  &  -  \\
 J022345.04-054234.4  &   0.143  &    9.1$\pm$0.5&   58.7$\pm$2.3  &   22.7$\pm$2.2  &  -  \\
 J022413.64-042227.8  &   0.116  &    9.2$\pm$0.5&   55.3$\pm$2.3  &   25.8$\pm$2.6  &  -  \\
 J022422.48-040230.5  &   0.171  &    7.5$\pm$0.4&   31.4$\pm$2.4  &  -  &  -  \\
 J022431.58-052818.8  &   2.068  &    9.4$\pm$0.5&   45.7$\pm$2.4  &   42.0$\pm$2.3  &   28.1$\pm$3.5  \\
 J022434.28-041531.2  &   0.259  &    6.3$\pm$0.3&   73.1$\pm$2.4  &   21.8$\pm$3.5  &  -  \\
 J022438.97-042706.3  &   0.252  &    6.6$\pm$0.3&   31.4$\pm$2.4  &   27.8$\pm$2.4  &   24.6$\pm$3.0  \\
 J022446.99-040851.3  &   0.096  &    5.3$\pm$0.3&  131.8$\pm$2.4  &   41.5$\pm$3.9  &  -  \\
 J022457.64-041417.9  &   0.063  &   11.9$\pm$0.6&  166.6$\pm$2.3  &   91.5$\pm$2.5  &   33.3$\pm$6.1  \\
 J022507.43-041835.7  &   0.105  &    6.8$\pm$0.3&   66.1$\pm$2.2  &   31.4$\pm$2.3  &   20.4$\pm$2.6  \\
 J022522.59-045452.2  &   0.144  &   10.1$\pm$0.5&  115.1$\pm$2.4  &   53.1$\pm$3.7  &  -  \\
 J022536.44-050011.5  &   0.053  &   13.7$\pm$0.7&  474.6$\pm$2.5  &  205.1$\pm$2.6  &   75.1$\pm$3.6  \\
 J022548.21-050051.5  &   0.150  &    8.0$\pm$0.4&   59.1$\pm$2.3  &  -  &  -  \\
 J022549.78-040024.6  &   0.044  &   58.5$\pm$2.9&  186.1$\pm$2.5  &   73.5$\pm$2.7  &   30.0$\pm$3.2  \\
 J022559.99-050145.3  &   0.205  &    5.7$\pm$0.3&   68.4$\pm$2.6  &   30.3$\pm$2.3  &  -  \\
 J022602.92-045306.8  &   0.056  &    6.4$\pm$0.3&   59.3$\pm$2.4  &   31.4$\pm$2.5  &   24.8$\pm$2.7  \\
 J022603.61-045903.8  &   0.055  &   31.4$\pm$1.6&  167.6$\pm$2.5  &   75.7$\pm$2.4  &   30.2$\pm$3.1  \\
 J022617.43-050443.4  &   0.057  &   48.7$\pm$2.4&   79.9$\pm$2.5  &   35.5$\pm$2.5  &  -  \\
 J022637.79-035841.6  &   0.070  &   13.5$\pm$0.7&   28.4$\pm$2.5  &   18.4$\pm$4.0  &  -  \\
 J022655.87-040302.2  &   0.135  &    6.9$\pm$0.3&   16.1$\pm$2.4  &  -  &  -  \\
 J022720.68-044537.1  &   0.055  &   73.1$\pm$3.7&  329.5$\pm$2.5  &  140.4$\pm$2.4  &   52.0$\pm$3.5  \\
 J022738.53-044702.7  &   0.173  &    7.1$\pm$0.4&   25.6$\pm$2.4  &  -  &  -  \\
 J022741.64-045650.5  &   0.055  &   11.4$\pm$0.6&  126.1$\pm$2.5  &   56.6$\pm$2.6  &   31.8$\pm$3.0  \\
 J103237.44+580845.9  &   0.251  &    6.1$\pm$0.3&  169.7$\pm$8.0  &   62.3$\pm$3.5  &  -  \\
 J103450.50+584418.2  &   0.091  &   20.1$\pm$1.0&   99.0$\pm$6.4  &   44.2$\pm$9.5  &  -  \\
 J103513.72+573444.6  &   1.537  &    5.5$\pm$0.3&   69.5$\pm$4.3  &   45.3$\pm$3.6  &   33.0$\pm$5.4  \\
 J103527.20+583711.9  &   0.885  &    6.9$\pm$0.3&   52.8$\pm$4.1  &   42.5$\pm$4.1  &   21.3$\pm$6.6  \\
 J103531.46+581234.2  &   0.176  &    5.0$\pm$0.3&  108.6$\pm$5.1  &   43.5$\pm$4.8  &  -  \\
 J103542.76+583313.1  &   0.087  &    6.6$\pm$0.3&   27.7$\pm$4.4  &  -  &  -  \\
 J103601.81+581836.2  &   0.100  &    6.0$\pm$0.3&   53.3$\pm$3.9  &   21.3$\pm$4.5  &  -  \\
 J103646.42+584330.6  &   0.140  &    6.8$\pm$0.3&   60.2$\pm$3.9  &   18.0$\pm$4.2  &  -  \\
 J103803.35+572701.5  &   1.285  &   15.4$\pm$0.8&   70.5$\pm$4.3  &   37.8$\pm$4.3  &  -  \\
 J103813.90+580047.3  &   0.205  &    6.2$\pm$0.3&   30.8$\pm$4.4  &   17.0$\pm$4.4  &  -  \\
 J103856.16+570333.9  &   0.178  &    5.7$\pm$0.3&   16.6$\pm$4.3  &  -  &  -  \\
 J104016.32+570846.0  &   0.118  &    5.2$\pm$0.3&   63.6$\pm$5.6  &  -  &  -  \\
 J104058.79+581703.3  &   0.072  &   10.4$\pm$0.5&   16.8$\pm$4.1  &  -  &  -  \\
 J104131.79+592258.4  &   0.925  &    7.0$\pm$0.4&   41.0$\pm$4.3  &  -  &  -  \\
 J104132.49+565953.0  &   0.346  &    8.3$\pm$0.4&   21.0$\pm$4.0  &  -  &  -  \\
 J104159.83+585856.4  &   0.360  &   21.7$\pm$1.1&   31.5$\pm$4.9  &   17.7$\pm$5.3  &  -  \\
 J104255.66+575549.7  &   1.468  &    6.4$\pm$0.3&   23.3$\pm$4.1  &  -  &  -  \\
 J104432.94+564041.6  &   0.067  &   28.7$\pm$1.4&  286.5$\pm$5.1  &  114.4$\pm$4.4  &   31.8$\pm$6.9  \\
 J104438.21+562210.7  &   0.025  &   80.6$\pm$4.0&  590.6$\pm$7.4  &  231.0$\pm$5.3  &   84.6$\pm$5.6  \\
 J104454.08+574425.7  &   0.118  &    6.5$\pm$0.3&  165.3$\pm$4.9  &   77.3$\pm$4.7  &  -  \\
 J104516.02+592304.7  &   0.322  &    5.1$\pm$0.3&   19.4$\pm$4.1  &  -  &  -  \\
 J104643.26+584715.1  &   0.140  &    5.4$\pm$0.3&  102.4$\pm$4.5  &   42.1$\pm$5.6  &  -  \\
 J104705.07+590728.4  &   0.391  &    7.0$\pm$0.4&   16.1$\pm$4.3  &  -  &  -  \\
 J104729.89+572842.9  &   0.230  &    6.2$\pm$0.3&  136.0$\pm$4.8  &   72.8$\pm$4.5  &   26.6$\pm$7.5  \\
 J104837.81+582642.1  &   0.232  &    7.6$\pm$0.4&  114.0$\pm$4.3  &   46.5$\pm$4.8  &  -  \\
 J104843.90+580341.2  &   0.162  &    7.1$\pm$0.4&   36.8$\pm$6.5  &  -  &  -  \\
 J104907.15+565715.3  &   0.072  &    9.7$\pm$0.5&  106.9$\pm$4.1  &   31.8$\pm$4.4  &  -  \\
 J105005.97+561500.0  &   0.119  &   14.8$\pm$0.7&  106.9$\pm$3.7  &   44.0$\pm$3.8  &   27.9$\pm$7.4  \\
 J105047.83+590348.3  &   0.131  &    5.2$\pm$0.3&   75.1$\pm$5.4  &   32.4$\pm$4.1  &  -  \\
 J105106.12+591625.3  &   0.768  &    5.4$\pm$0.3&   29.8$\pm$6.5  &  -  &  -  \\
 J105128.05+573502.4  &   0.073  &   10.0$\pm$0.5&   54.6$\pm$4.4  &   24.4$\pm$4.4  &  -  \\
 J105158.53+590652.0  &   1.814  &    5.4$\pm$0.3&   48.7$\pm$5.2  &   40.7$\pm$8.4  &  -  \\
 J105200.29+591933.7  &   0.115  &   11.4$\pm$0.6&   31.3$\pm$5.1  &  -  &  -  \\
 J105206.56+580947.1  &   0.117  &   16.7$\pm$0.8&  247.5$\pm$6.0  &   97.7$\pm$5.8  &  -  \\
 J105336.87+580350.7  &   0.460  &    5.9$\pm$0.3&   33.3$\pm$4.7  &  -  &  -  \\
 J105404.11+574019.7  &   1.101  &    8.5$\pm$0.4&   15.5$\pm$4.2  &  -  &  -  \\
 J105421.65+582344.6  &   0.205  &   16.8$\pm$0.8&  114.7$\pm$4.1  &   39.9$\pm$3.8  &  -  \\
 J105604.84+574229.9  &   1.211  &   11.2$\pm$0.6&   36.6$\pm$6.9  &   28.5$\pm$8.7  &  -  \\
 J105636.95+573449.3  &   0.047  &    6.4$\pm$0.3&  109.9$\pm$4.5  &   35.6$\pm$5.6  &  -  \\
 J105641.81+580046.0  &   0.130  &    7.5$\pm$0.4&  104.9$\pm$4.2  &   34.1$\pm$4.0  &  -  \\
 J105705.43+580437.4  &   0.140  &   16.5$\pm$0.8&  129.1$\pm$4.6  &   47.4$\pm$4.7  &  -  \\
 J105733.53+565737.4  &   0.086  &    5.6$\pm$0.3&   26.1$\pm$5.3  &  -  &  -  \\
 J105740.55+570616.4  &   0.073  &    6.1$\pm$0.3&   34.7$\pm$3.9  &  -  &  -  \\
 J105903.47+572155.1  &   0.119  &   13.8$\pm$0.7&   60.9$\pm$4.5  &  -  &  -  \\
 J105951.71+581802.9  &   2.335  &    5.3$\pm$0.3&   29.7$\pm$4.6  &   41.8$\pm$4.8  &   27.0$\pm$8.8  \\
 J105959.95+574848.1  &   0.453  &    9.1$\pm$0.5&   37.8$\pm$5.2  &  -  &  -  \\
 J110002.06+573142.1  &   0.387  &    8.3$\pm$0.4&   64.0$\pm$4.3  &  -  &  -  \\
 J110124.97+574315.8  &   0.243  &    6.1$\pm$0.3&   47.1$\pm$4.1  &   25.2$\pm$4.7  &  -  \\
 J110133.80+575206.6  &   0.277  &    6.4$\pm$0.3&  136.3$\pm$5.2  &   58.2$\pm$4.5  &   30.6$\pm$6.5  \\
 J110223.58+574436.2  &   0.226  &   10.2$\pm$0.5&   27.0$\pm$5.1  &   17.4$\pm$5.5  &  -  \\
 J110235.02+574655.7  &   0.226  &    6.2$\pm$0.3&   44.2$\pm$4.3  &  -  &  -  \\
 J160408.18+542531.2  &   0.260  &    5.0$\pm$0.3&   62.4$\pm$1.5  &   33.1$\pm$1.6  &  -  \\
 J160655.35+534016.9  &   0.214  &   14.6$\pm$0.7&   30.1$\pm$1.6  &   23.1$\pm$1.5  &  -  \\
 J160803.71+545301.9  &   0.053  &    5.1$\pm$0.3&  180.7$\pm$1.5  &   76.2$\pm$1.6  &   25.8$\pm$2.6  \\
 J160832.59+552926.9  &   0.065  &    5.9$\pm$0.3&  102.9$\pm$1.8  &   50.6$\pm$1.6  &  -  \\
 J160858.38+553010.2  &   0.066  &    8.8$\pm$0.4&  101.7$\pm$1.3  &   37.2$\pm$1.2  &  -  \\
 J160907.56+552428.4  &   0.065  &    7.7$\pm$0.4&  128.1$\pm$1.5  &   57.9$\pm$1.5  &   19.1$\pm$2.0  \\
 J160908.28+552241.4  &   0.084  &    6.6$\pm$0.3&  110.9$\pm$1.5  &   42.5$\pm$1.7  &  -  \\
 J160926.69+551642.3  &   0.068  &    6.8$\pm$0.3&   81.7$\pm$1.5  &   35.8$\pm$1.9  &  -  \\
 J160931.55+541827.3  &   0.082  &    5.6$\pm$0.3&  105.1$\pm$1.5  &   35.0$\pm$1.5  &  -  \\
 J160937.48+541259.2  &   0.086  &    5.7$\pm$0.3&  129.7$\pm$1.5  &   58.0$\pm$1.5  &   22.1$\pm$2.0  \\
 J161103.73+544322.0  &   0.063  &    6.6$\pm$0.3&   71.4$\pm$1.5  &   29.4$\pm$3.3  &  -  \\
 J161123.44+545158.2  &   0.078  &    5.5$\pm$0.3&   56.2$\pm$1.6  &   21.4$\pm$1.5  &  -  \\
 J161223.39+540339.2  &   0.138  &   13.0$\pm$0.6&   70.5$\pm$1.6  &   28.1$\pm$1.9  &  -  \\
 J161233.43+545630.4  &   0.083  &    8.3$\pm$0.4&  185.3$\pm$1.5  &   87.0$\pm$1.8  &   38.3$\pm$3.5  \\
 J161241.05+543956.8  &   0.035  &    5.7$\pm$0.3&   22.0$\pm$1.5  &  -  &  -  \\
 J161250.85+532304.9  &   0.048  &   17.9$\pm$0.9&  114.6$\pm$1.6  &   46.3$\pm$1.2  &   18.6$\pm$1.9  \\
 J161254.17+545525.4  &   0.065  &    8.0$\pm$0.4&  257.0$\pm$1.6  &  123.6$\pm$1.5  &   49.4$\pm$1.8  \\
 J161357.01+534105.3  &   0.180  &    6.5$\pm$0.3&   45.9$\pm$1.6  &   32.9$\pm$1.8  &   22.9$\pm$2.1  \\
 J161411.52+540554.3  &   0.305  &    5.9$\pm$0.3&   47.7$\pm$1.6  &   28.5$\pm$1.8  &   19.2$\pm$1.9  \\
 J161521.78+543148.3  &   0.474  &    5.1$\pm$0.3&   21.7$\pm$1.5  &  -  &  -  \\
 J161551.45+541535.9  &   0.215  &    6.3$\pm$0.3&   98.7$\pm$1.5  &   42.0$\pm$1.5  &  -  \\
 J161645.92+542554.4  &   0.223  &   12.4$\pm$0.6&   31.4$\pm$1.6  &  -  &  -  \\
 J161759.22+541501.3  &   0.135  &   22.7$\pm$1.1&   35.8$\pm$2.4  &   16.5$\pm$3.1  &  -  \\
 J161819.31+541859.0  &   0.083  &   28.3$\pm$1.4&  378.9$\pm$1.7  &  153.5$\pm$1.9  &   51.4$\pm$3.1  \\
 J171033.21+584456.8  &   0.281  &    6.1$\pm$0.3&   43.7$\pm$2.8  &  -  &  -  \\
 J171232.34+592125.9  &   0.210  &    8.7$\pm$0.4&  138.9$\pm$3.0  &   85.3$\pm$3.2  &   36.9$\pm$4.2  \\
 J171233.38+583610.5  &   1.663  &    5.1$\pm$0.3&   20.7$\pm$2.5  &  -  &  -  \\
 J171233.77+594026.4  &   0.217  &    5.1$\pm$0.3&   36.4$\pm$3.4  &  -  &  -  \\
 J171316.50+583234.9  &   0.079  &    6.7$\pm$0.3&  139.5$\pm$3.1  &   60.7$\pm$3.0  &   21.5$\pm$3.6  \\
 J171414.81+585221.5  &   0.167  &    9.0$\pm$0.5&   38.5$\pm$2.5  &  -  &  -  \\
 J171419.98+602724.6  &   2.990  &    5.6$\pm$0.3&   17.9$\pm$2.5  &  -  &  -  \\
 J171446.47+593400.1  &   0.129  &    7.5$\pm$0.4&   96.9$\pm$2.8  &   34.3$\pm$3.1  &  -  \\
 J171447.31+583805.9  &   0.257  &    5.4$\pm$0.3&   61.3$\pm$2.7  &   27.5$\pm$2.3  &  -  \\
 J171513.88+594638.1  &   0.248  &    5.1$\pm$0.3&   60.6$\pm$2.6  &   25.1$\pm$2.8  &  -  \\
 J171550.50+593548.8  &   0.066  &    9.1$\pm$0.5&  123.7$\pm$2.8  &   63.2$\pm$2.7  &   25.5$\pm$3.6  \\
 J171614.48+595423.8  &   0.153  &    8.6$\pm$0.4&  107.5$\pm$2.9  &   45.6$\pm$2.8  &  -  \\
 J171630.23+601422.7  &   0.107  &    8.6$\pm$0.4&   41.7$\pm$2.5  &  -  &  -  \\
 J171650.58+595751.4  &   0.182  &    6.8$\pm$0.3&   17.0$\pm$3.2  &  -  &  -  \\
 J171711.11+602710.0  &   0.110  &    9.5$\pm$0.5&   58.7$\pm$2.4  &   20.0$\pm$2.6  &  -  \\
 J171747.51+593258.1  &   0.248  &    5.3$\pm$0.3&   22.3$\pm$2.4  &  -  &  -  \\
 J171852.71+591432.0  &   0.322  &   14.0$\pm$0.7&   51.4$\pm$2.6  &   24.7$\pm$2.8  &  -  \\
 J171933.37+592742.8  &   0.139  &    7.6$\pm$0.4&  128.3$\pm$2.8  &   45.4$\pm$2.7  &  -  \\
 J171944.91+595707.7  &   0.069  &   14.4$\pm$0.7&  135.3$\pm$3.2  &   50.4$\pm$3.1  &  -  \\
 J172043.28+584026.6  &   0.125  &    9.7$\pm$0.5&  144.8$\pm$2.8  &   50.8$\pm$2.6  &  -  \\
 J172044.86+582924.0  &   1.697  &    5.3$\pm$0.3&   24.1$\pm$2.7  &  -  &  -  \\
 J172159.43+595034.3  &   0.028  &    9.7$\pm$0.5&  146.5$\pm$2.7  &   62.2$\pm$2.9  &  -  \\
 J172219.58+594506.9  &   0.272  &    7.8$\pm$0.4&   25.2$\pm$3.2  &   19.0$\pm$3.3  &  -  \\
 J172228.04+601526.0  &   0.742  &    7.2$\pm$0.4&   40.1$\pm$3.0  &   18.3$\pm$3.8  &  -  \\
 J172238.73+585107.0  &   1.624  &    6.7$\pm$0.3&   54.1$\pm$2.5  &   42.8$\pm$2.9  &   29.0$\pm$3.4  \\
 J172313.06+590533.1  &   0.108  &    6.2$\pm$0.3&   61.1$\pm$2.4  &   24.4$\pm$2.8  &  -  \\
 J172355.58+601301.7  &   0.175  &    5.4$\pm$0.3&   89.4$\pm$2.6  &   36.8$\pm$2.7  &   19.3$\pm$3.4  \\
 J172355.97+594047.6  &   0.030  &    5.2$\pm$0.3&   27.2$\pm$2.4  &  -  &  -  \\
 J172402.11+600601.4  &   0.156  &    8.0$\pm$0.4&   70.7$\pm$5.0  &  -  &  -  \\
 J172546.80+593655.3  &   0.035  &   26.0$\pm$1.3&  461.0$\pm$4.5  &  201.7$\pm$2.9  &   83.6$\pm$3.9  \\
 J172551.35+601138.9  &   0.029  &   27.3$\pm$1.4&  266.7$\pm$3.4  &  115.8$\pm$3.0  &   44.0$\pm$2.9  \\
 \enddata
\end{deluxetable}

\begin{deluxetable}{lccccccc}

\tabletypesize{\normalsize}
\tablewidth{0pc}
\tablecaption{Physical Properties of the 5MUSES sample as derived based on the DL07 and MBB models}
\tablehead{

\colhead{Name} &
\colhead{$z_{\rm spec}$} &
\colhead{log$L_{\rm IR}$} &
\colhead{log$M_{\rm dust}$ $^{a}$} &
\colhead{$\gamma$} &
\colhead{$U_{min}$} &
\colhead{$T_{\rm d}$ $^{b}$}&
\colhead{AGN $^{c}$} \\

\colhead{} &
\colhead{} &
\colhead{[$L_{\odot}$]} &
\colhead{[$M_{\odot}$]}&
\colhead{-}&
\colhead{-}&
\colhead{[K]}&
\colhead{frac.}} 
\startdata
 J021503.52-042421.6  &   0.137  &  10.86$\pm$0.04&   8.01$\pm$0.07  &   1.0&   5&  -  &   0.06  \\
 J021557.11-033729.0  &   0.032  &   9.83$\pm$0.02&   7.18$\pm$0.09  &   1.5&   3&   26$\pm$1  &   0.00  \\
 J021638.21-042250.8  &   0.143  &  10.99$\pm$0.01&  -  &   -& -&  -  &   0.58  \\
 J021640.72-044405.1  &   0.031  &  10.11$\pm$0.01&   7.29$\pm$0.07  &   0.7&   5&   28$\pm$1  &   0.08  \\
 J021649.71-042554.8  &   0.226  &  11.67$\pm$0.01&   8.76$\pm$0.03  &   3.0&   5&   31$\pm$2  &   0.36  \\
 J021657.77-032459.7  &   0.292  &  11.64$\pm$0.03&   8.35$\pm$0.02  &   0.7&  12&   35$\pm$1  &   0.16  \\
 J021729.06-041937.8  &   0.199  &  11.13$\pm$0.01&  -  &   -& -&  -  &   0.12  \\
 J021743.01-043625.1  &   0.103  &  10.66$\pm$0.01&  -  &   -& -&   30$\pm$1  &   0.74  \\
 J021743.82-051751.7  &   0.042  &   9.98$\pm$0.01&   6.50$\pm$0.06  &  10.0&  10&   35$\pm$7  &   0.29  \\
 J021754.88-035826.4  &   0.151  &  11.36$\pm$0.01&  -  &   -& -&   36$\pm$1  &   0.58  \\
 J021808.22-045845.3  &   0.072  &  10.86$\pm$0.04&   7.75$\pm$0.03  &   3.0&   8&   32$\pm$1  &   0.12  \\
 J021830.57-045622.9  &   0.081  &  10.46$\pm$0.01&   7.65$\pm$0.04  &   0.7&   5&   29$\pm$1  &   0.08  \\
 J021849.76-052158.2  &   1.560  &  13.07$\pm$0.01&  -  &   -& -&  -  &   0.52  \\
 J021859.74-040237.2  &   0.008  &   8.18$\pm$0.01&   5.43$\pm$0.02  &   3.5&   3&   27$\pm$1  &   0.38  \\
 J021909.60-052512.9  &   0.068  &  10.96$\pm$0.04&   8.09$\pm$0.03  &   1.5&   5&   29$\pm$1  &   0.10  \\
 J021912.71-050541.8  &   0.025  &   9.23$\pm$0.02&   6.52$\pm$0.07  &   2.5&   3&  -  &   0.30  \\
 J021916.05-055726.9  &   0.164  &  11.10$\pm$0.02&   7.91$\pm$0.08  &   5.0&   8&   33$\pm$1  &   0.00  \\
 J021928.33-042239.8  &   0.258  &  11.67$\pm$0.03&   8.50$\pm$0.09  &   0.1&  10&   33$\pm$2  &   0.18  \\
 J021938.70-032508.2  &   0.073  &  10.26$\pm$0.02&   7.33$\pm$0.07  &   3.5&   5&   30$\pm$1  &   0.00  \\
 J021939.08-051133.8  &   0.149  &  10.93$\pm$0.01&   7.55$\pm$0.04  &   3.0&  12&   35$\pm$1  &   0.16  \\
 J021953.04-051824.1  &   0.139  &  10.51$\pm$0.01&   7.83$\pm$0.10  &   8.0&   2&  -  &   0.55  \\
 J021956.96-052440.4  &   0.239  &  11.25$\pm$0.02&   8.32$\pm$0.04  &   4.0&   5&   31$\pm$1  &   0.40  \\
 J022000.22-043947.6  &   0.708  &  12.67$\pm$0.01&  -  &   -& -&   43$\pm$1  &   0.40  \\
 J022005.93-031545.7  &   0.199  &  11.31$\pm$0.02&   8.33$\pm$0.07  &   4.5&   5&   32$\pm$1  &   0.01  \\
 J022012.21-034111.8  &   0.144  &  11.00$\pm$0.01&   7.87$\pm$0.03  &   4.5&   7&   33$\pm$1  &   0.04  \\
 J022145.09-053207.4  &   0.176  &  11.02$\pm$0.01&   7.43$\pm$0.07  &   5.5&  15&  -  &   0.02  \\
 J022147.82-025730.7  &   0.143  &  11.08$\pm$0.03&   7.88$\pm$0.07  &   1.0&  10&  -  &   0.19  \\
 J022147.87-044613.5  &   0.116  &  10.90$\pm$0.02&   7.77$\pm$0.06  &   3.5&   8&   31$\pm$1  &   0.14  \\
 J022151.54-032911.8  &   0.171  &  11.09$\pm$0.01&   7.58$\pm$0.04  &   6.5&  12&  -  &   0.23  \\
 J022205.03-050537.0  &   2.068  &  13.35$\pm$0.01&  -  &   -& -&   30$\pm$2  &   0.54  \\
 J022223.26-044319.8  &   0.259  &  11.57$\pm$0.02&   8.29$\pm$0.07  &   3.0&  10&   34$\pm$7  &   0.18  \\
 J022224.06-050550.3  &   0.252  &  11.21$\pm$0.02&  -  &   -& -&   32$\pm$1  &   0.19  \\
 J022241.34-045652.0  &   0.096  &  10.89$\pm$0.02&   8.10$\pm$0.05  &   0.0&   5&   29$\pm$1  &   0.00  \\
 J022257.96-041840.8  &   0.063  &  10.69$\pm$0.04&   7.80$\pm$0.03  &   2.0&   5&   29$\pm$1  &   0.06  \\
 J022301.97-052335.8  &   0.105  &  10.61$\pm$0.02&   7.97$\pm$0.06  &   6.0&   2&   28$\pm$1  &   0.09  \\
 J022309.31-052316.1  &   0.144  &  11.27$\pm$0.01&   8.22$\pm$0.07  &   2.5&   7&   31$\pm$1  &   0.00  \\
 J022315.58-040606.0  &   0.053  &  11.02$\pm$0.05&   8.00$\pm$0.05  &   0.4&   8&   31$\pm$1  &   0.08  \\
 J022329.13-043209.5  &   0.150  &  11.15$\pm$0.01&   7.72$\pm$0.07  &   0.5&  15&   36$\pm$1  &   0.12  \\
 J022334.65-035229.4  &   0.044  &  10.62$\pm$0.01&   7.39$\pm$0.02  &   6.5&   8&   32$\pm$1  &   0.23  \\
 J022345.04-054234.4  &   0.205  &  11.28$\pm$0.03&   8.27$\pm$0.06  &   2.0&   7&   32$\pm$6  &   0.17  \\
 J022356.49-025431.1  &   0.056  &  10.13$\pm$0.02&   7.25$\pm$0.10  &   2.0&   5&   29$\pm$1  &   0.09  \\
 J022413.64-042227.8  &   0.055  &  10.59$\pm$0.03&   7.68$\pm$0.05  &   2.5&   5&   29$\pm$6  &   0.10  \\
 J022422.48-040230.5  &   0.057  &  10.72$\pm$0.01&  -  &   -& -&   38$\pm$1  &   0.68  \\
 J022431.58-052818.8  &   0.070  &  10.42$\pm$0.03&   7.26$\pm$0.07  &   4.0&   8&  -  &   0.11  \\
 J022434.28-041531.2  &   0.135  &  10.66$\pm$0.01&  -  &   -& -&  -  &   0.26  \\
 J022438.97-042706.3  &   0.055  &  11.02$\pm$0.01&   7.86$\pm$0.03  &   4.0&   8&   32$\pm$1  &   0.16  \\
 J022446.99-040851.3  &   0.173  &  11.07$\pm$0.01&   7.51$\pm$0.03  &   2.0&  20&  -  &   0.21  \\
 J022457.64-041417.9  &   0.055  &  10.51$\pm$0.02&   7.79$\pm$0.07  &   0.7&   4&   27$\pm$1  &   0.10  \\
 J022507.43-041835.7  &   0.251  &  11.80$\pm$0.01&   8.82$\pm$0.09  &   0.7&   7&   31$\pm$1  &   0.00  \\
 J022508.33-053917.7  &   0.091  &  10.85$\pm$0.01&   7.86$\pm$0.07  &   5.5&   5&   31$\pm$1  &   0.14  \\
 J022522.59-045452.2  &   1.537  &  13.13$\pm$0.01&  -  &   -& -&   46$\pm$2  &   0.74  \\
 J022536.44-050011.5  &   0.885  &  12.57$\pm$0.01&  -  &   -& -&   35$\pm$1  &   0.42  \\
 J022548.21-050051.5  &   0.176  &  11.27$\pm$0.01&   8.27$\pm$0.05  &   1.5&   7&   31$\pm$1  &   0.01  \\
 J022549.78-040024.6  &   0.087  &  10.48$\pm$0.01&   7.12$\pm$0.08  &   2.5&  12&  -  &   0.00  \\
 J022559.99-050145.3  &   0.100  &  10.54$\pm$0.02&   7.86$\pm$0.07  &   2.5&   3&   28$\pm$2  &   0.13  \\
 J022602.92-045306.8  &   0.140  &  10.94$\pm$0.01&   7.88$\pm$0.05  &   3.0&   7&   32$\pm$1  &   0.00  \\
 J022603.61-045903.8  &   1.285  &  13.20$\pm$0.01&  -  &   -& -&   54$\pm$1  &   0.78  \\
 J022617.43-050443.4  &   0.205  &  11.12$\pm$0.01&   7.83$\pm$0.03  &   8.0&   8&  -  &   0.36  \\
 J022637.79-035841.6  &   0.178  &  10.74$\pm$0.02&   6.56$\pm$3.71  &  90.0&  10&  -  &   0.24  \\
 J022655.87-040302.2  &   0.118  &  10.80$\pm$0.01&   7.73$\pm$0.04  &   2.0&   8&  -  &   0.08  \\
 J022720.68-044537.1  &   0.072  &   9.91$\pm$0.01&  -  &   -& -&  -  &   0.56  \\
 J022738.53-044702.7  &   0.925  &  12.32$\pm$0.01&  -  &   -& -&  -  &   0.66  \\
 J022741.64-045650.5  &   0.346  &  11.75$\pm$0.01&   7.94$\pm$0.03  &  10.0&  20&  -  &   0.13  \\
 J103237.44+580845.9  &   0.360  &  11.91$\pm$0.01&  -  &   -& -&  -  &   0.52  \\
 J103450.50+584418.2  &   1.468  &  12.81$\pm$0.01&  -  &   -& -&  -  &   0.64  \\
 J103513.72+573444.6  &   0.067  &  10.94$\pm$0.02&   8.04$\pm$0.06  &   2.0&   5&   30$\pm$7  &   0.00  \\
 J103527.20+583711.9  &   0.025  &  10.44$\pm$0.04&   7.46$\pm$0.04  &   1.0&   7&   30$\pm$6  &   0.11  \\
 J103531.46+581234.2  &   0.118  &  11.00$\pm$0.07&   8.39$\pm$0.10  &   0.9&   3&   26$\pm$1  &   0.14  \\
 J103542.76+583313.1  &   0.322  &  11.44$\pm$0.01&  -  &   -& -&  -  &   0.42  \\
 J103601.81+581836.2  &   0.140  &  10.90$\pm$0.02&   8.37$\pm$0.07  &   3.0&   2&   27$\pm$1  &   0.15  \\
 J103606.45+581829.7  &   0.391  &  11.58$\pm$0.01&  -  &   -& -&  -  &   0.95  \\
 J103646.42+584330.6  &   0.230  &  11.54$\pm$0.02&   8.75$\pm$0.06  &   0.4&   5&   29$\pm$1  &   0.11  \\
 J103701.99+574414.8  &   0.232  &  11.66$\pm$0.02&   8.66$\pm$0.08  &   1.5&   7&  -  &   0.10  \\
 J103724.74+580512.9  &   0.162  &  11.02$\pm$0.01&   7.68$\pm$0.05  &   5.0&  10&  -  &   0.00  \\
 J103803.35+572701.5  &   0.072  &  10.62$\pm$0.01&   7.60$\pm$0.07  &   0.9&   8&  -  &   0.07  \\
 J103813.90+580047.3  &   0.119  &  11.13$\pm$0.03&   7.77$\pm$0.06  &   5.0&  10&   34$\pm$1  &   0.09  \\
 J103818.19+583556.5  &   0.131  &  10.86$\pm$0.04&   8.12$\pm$0.05  &   0.9&   4&   29$\pm$1  &   0.00  \\
 J103856.16+570333.9  &   0.768  &  12.23$\pm$0.01&  -  &   -& -&  -  &   0.39  \\
 J104016.32+570846.0  &   0.073  &  10.40$\pm$0.01&   7.06$\pm$0.06  &   4.5&  10&   34$\pm$1  &   0.00  \\
 J104058.79+581703.3  &   1.814  &  13.16$\pm$0.01&  -  &   -& -&  -  &   0.51  \\
 J104131.79+592258.4  &   0.115  &  10.78$\pm$0.01&   7.28$\pm$0.08  &   6.5&  12&   36$\pm$1  &   0.13  \\
 J104132.49+565953.0  &   0.117  &  11.31$\pm$0.01&   8.49$\pm$0.06  &   0.7&   5&   29$\pm$1  &   0.12  \\
 J104159.83+585856.4  &   0.460  &  11.92$\pm$0.01&   8.99$\pm$0.21  &  20.0&   2&  -  &   0.08  \\
 J104255.66+575549.7  &   1.101  &  12.50$\pm$0.02&  -  &   -& -&  -  &   0.83  \\
 J104303.50+585718.1  &   0.205  &  11.47$\pm$0.01&  -  &   -& -&   32$\pm$1  &   0.37  \\
 J104432.94+564041.6  &   1.211  &  12.98$\pm$0.02&  -  &   -& -&   49$\pm$1  &   0.78  \\
 J104438.21+562210.7  &   0.047  &  10.22$\pm$0.01&   7.25$\pm$0.09  &   0.4&   7&   31$\pm$1  &   0.00  \\
 J104454.08+574425.7  &   0.130  &  11.00$\pm$0.02&   7.98$\pm$0.09  &   2.0&   7&   30$\pm$6  &   0.00  \\
 J104501.73+571111.3  &   0.140  &  11.19$\pm$0.02&  -  &   -& -&   30$\pm$1  &   0.58  \\
 J104516.02+592304.7  &   0.086  &  10.32$\pm$0.01&   6.83$\pm$0.07  &   2.5&  15&  -  &   0.08  \\
 J104643.26+584715.1  &   0.073  &  10.26$\pm$0.01&   7.31$\pm$0.07  &   4.0&   5&  -  &   0.00  \\
 J104705.07+590728.4  &   0.119  &  10.87$\pm$0.02&   7.54$\pm$0.09  &  10.0&   8&   33$\pm$1  &   0.31  \\
 J104729.89+572842.9  &   2.335  &  13.40$\pm$0.02&  -  &   -& -&   30$\pm$1  &   0.62  \\
 J104837.81+582642.1  &   0.453  &  11.78$\pm$0.01&  -  &   -& -&  -  &   0.38  \\
 J104839.73+555356.4  &   0.387  &  12.02$\pm$0.01&   8.26$\pm$0.02  &   8.0&  20&  -  &   0.07  \\
 J104843.90+580341.2  &   0.243  &  11.26$\pm$0.01&   8.22$\pm$0.06  &   6.5&   5&   33$\pm$1  &   0.08  \\
 J104907.15+565715.3  &   0.277  &  11.81$\pm$0.01&   8.75$\pm$0.04  &   1.5&   8&   33$\pm$2  &   0.00  \\
 J104918.33+562512.9  &   0.226  &  11.25$\pm$0.01&  -  &   -& -&   29$\pm$1  &   0.36  \\
 J105005.97+561500.0  &   0.226  &  11.46$\pm$0.01&   7.83$\pm$0.02  &   3.5&  20&  -  &   0.08  \\
 J105047.83+590348.3  &   0.260  &  11.52$\pm$0.02&   8.29$\pm$0.04  &   1.5&  10&   34$\pm$2  &   0.07  \\
 J105058.76+560550.0  &   0.214  &  11.26$\pm$0.02&  -  &   -& -&   34$\pm$1  &   0.20  \\
 J105106.12+591625.3  &   0.053  &  10.35$\pm$0.04&   7.78$\pm$0.05  &   0.0&   3&   26$\pm$1  &   0.00  \\
 J105128.05+573502.4  &   0.065  &  10.27$\pm$0.02&   7.58$\pm$0.03  &   3.0&   3&   28$\pm$3  &   0.01  \\
 J105158.53+590652.0  &   0.066  &  10.34$\pm$0.01&   7.52$\pm$0.02  &   2.5&   4&   29$\pm$1  &   0.00  \\
 J105200.29+591933.7  &   0.065  &  10.56$\pm$0.03&   7.76$\pm$0.03  &   0.2&   5&   29$\pm$2  &   0.00  \\
 J105206.56+580947.1  &   0.084  &  10.62$\pm$0.01&   7.80$\pm$0.03  &   1.0&   5&   28$\pm$1  &   0.00  \\
 J105336.87+580350.7  &   0.068  &  10.30$\pm$0.02&   7.53$\pm$0.04  &   1.5&   4&   29$\pm$1  &   0.00  \\
 J105404.11+574019.7  &   0.082  &  10.61$\pm$0.04&   7.81$\pm$0.05  &   0.2&   5&  -  &   0.00  \\
 J105421.65+582344.6  &   0.086  &  10.68$\pm$0.03&   7.96$\pm$0.04  &   0.6&   4&   28$\pm$1  &   0.00  \\
 J105604.84+574229.9  &   0.063  &  10.20$\pm$0.02&   7.56$\pm$0.03  &   1.5&   3&  -  &   0.16  \\
 J105636.95+573449.3  &   0.078  &  10.40$\pm$0.02&   7.36$\pm$0.05  &   1.0&   8&   31$\pm$1  &   0.00  \\
 J105641.81+580046.0  &   0.138  &  11.05$\pm$0.01&   7.78$\pm$0.03  &  10.0&   7&   34$\pm$3  &   0.24  \\
 J105705.43+580437.4  &   0.083  &  10.72$\pm$0.03&   8.22$\pm$0.02  &   2.0&   2&   26$\pm$1  &   0.00  \\
 J105733.53+565737.4  &   0.035  &   9.54$\pm$0.01&   6.44$\pm$0.04  &   2.5&   8&   31$\pm$1  &   0.00  \\
 J105740.55+570616.4  &   0.048  &  10.40$\pm$0.02&   7.20$\pm$0.02  &   0.9&  10&   33$\pm$1  &   0.00  \\
 J105829.28+580439.2  &   0.065  &  10.59$\pm$0.03&   8.17$\pm$0.06  &   0.6&   2&  -  &   0.00  \\
 J105854.08+574130.0  &   0.180  &  10.86$\pm$0.01&  -  &   -& -&   27$\pm$1  &   0.17  \\
 J105903.47+572155.1  &   0.305  &  11.73$\pm$0.03&   8.30$\pm$0.06  &   4.0&  12&   36$\pm$1  &   0.22  \\
 J105951.71+581802.9  &   0.474  &  11.59$\pm$0.01&  -  &   -& -&  -  &   0.43  \\
 J105959.95+574848.1  &   0.215  &  11.49$\pm$0.01&   8.36$\pm$0.03  &   3.0&   8&  -  &   0.00  \\
 J110002.06+573142.1  &   0.223  &  11.28$\pm$0.01&  -  &   -& -&  -  &   0.51  \\
 J110124.97+574315.8  &   0.135  &  11.15$\pm$0.01&  -  &   -& -&   38$\pm$2  &   0.67  \\
 J110133.80+575206.6  &   0.083  &  11.16$\pm$0.04&   8.31$\pm$0.03  &   0.9&   5&   29$\pm$1  &   0.00  \\
 J110223.58+574436.2  &   0.281  &  11.42$\pm$0.01&   8.22$\pm$0.02  &   6.5&   8&  -  &   0.21  \\
 J110235.02+574655.7  &   0.210  &  11.63$\pm$0.01&   8.55$\pm$0.03  &   1.5&   8&   32$\pm$2  &   0.00  \\
 J160408.18+542531.2  &   1.663  &  12.88$\pm$0.01&  -  &   -& -&  -  &   0.76  \\
 J160630.59+542007.4  &   0.217  &  11.26$\pm$0.01&   7.68$\pm$0.05  &   2.5&  20&  -  &   0.23  \\
 J160655.35+534016.9  &   0.079  &  10.59$\pm$0.01&   7.98$\pm$0.00  &   1.0&   3&   27$\pm$1  &   0.00  \\
 J160803.71+545301.9  &   0.167  &  11.19$\pm$0.01&   7.56$\pm$0.00  &   4.0&  20&  -  &   0.25  \\
 J160832.59+552926.9  &   2.990  &  13.14$\pm$0.02&  -  &   -& -&  -  &   0.00  \\
 J160839.73+552330.6  &   0.129  &  11.11$\pm$0.01&   8.05$\pm$0.07  &   2.0&   8&   32$\pm$1  &   0.00  \\
 J160858.38+553010.2  &   0.257  &  11.57$\pm$0.01&   8.18$\pm$0.01  &   3.0&  12&  -  &   0.34  \\
 J160907.56+552428.4  &   0.248  &  11.28$\pm$0.01&   8.45$\pm$0.07  &   3.5&   4&   31$\pm$9  &   0.04  \\
 J160908.28+552241.4  &   0.066  &  10.25$\pm$0.02&  -  &   -& -&   24$\pm$1  &   0.83  \\
 J160926.69+551642.3  &   0.153  &  11.25$\pm$0.02&   8.25$\pm$0.07  &   1.5&   7&   31$\pm$1  &   0.00  \\
 J160931.55+541827.3  &   0.107  &  10.62$\pm$0.05&   7.45$\pm$0.09  &   5.0&   8&  -  &   0.22  \\
 J160937.48+541259.2  &   0.182  &  10.81$\pm$0.01&   8.04$\pm$0.09  &  60.0&   0&  -  &   0.20  \\
 J161103.73+544322.0  &   0.110  &  10.82$\pm$0.01&   7.74$\pm$0.07  &   3.5&   7&   31$\pm$1  &   0.00  \\
 J161123.44+545158.2  &   0.248  &  10.84$\pm$0.01&  -  &   -& -&  -  &   0.56  \\
 J161223.39+540339.2  &   0.322  &  11.90$\pm$0.01&  -  &   -& -&  -  &   0.58  \\
 J161233.43+545630.4  &   0.139  &  11.22$\pm$0.02&   8.03$\pm$0.09  &   0.4&  10&   33$\pm$1  &   0.22  \\
 J161241.05+543956.8  &   0.069  &  10.73$\pm$0.01&   7.52$\pm$0.10  &   1.0&  10&  -  &   0.00  \\
 J161250.85+532304.9  &   0.125  &  11.11$\pm$0.02&   8.13$\pm$0.11  &   1.0&   7&   30$\pm$1  &   0.00  \\
 J161254.17+545525.4  &   1.697  &  12.91$\pm$0.01&  -  &   -& -&  -  &   0.56  \\
 J161357.01+534105.3  &   0.028  &   9.76$\pm$0.01&   7.07$\pm$0.07  &   0.4&   4&   27$\pm$1  &   0.00  \\
 J161411.52+540554.3  &   0.272  &  11.22$\pm$0.01&  -  &   -& -&  -  &   0.30  \\
 J161521.78+543148.3  &   0.742  &  12.34$\pm$0.01&  -  &   -& -&  -  &   0.55  \\
 J161551.45+541535.9  &   1.624  &  13.08$\pm$0.01&  -  &   -& -&  -  &   0.43  \\
 J161645.92+542554.4  &   0.108  &  10.74$\pm$0.02&   7.52$\pm$0.19  &   1.5&  10&   32$\pm$1  &   0.20  \\
 J161759.22+541501.3  &   0.175  &  11.15$\pm$0.01&   8.28$\pm$0.05  &   2.0&   5&   30$\pm$1  &   0.11  \\
 J161819.31+541859.0  &   0.030  &   9.33$\pm$0.01&   6.24$\pm$0.03  &   4.0&   7&  -  &   0.29  \\
 J171033.21+584456.8  &   0.156  &  11.10$\pm$0.01&   7.97$\pm$0.01  &   3.5&   8&  -  &   0.00  \\
 J171124.22+593121.4  &   0.035  &  10.47$\pm$0.02&   7.79$\pm$0.08  &   0.0&   4&   27$\pm$1  &   0.05  \\
 J171232.34+592125.9  &   0.029  &  10.25$\pm$0.01&   7.22$\pm$0.04  &   0.7&   8&   30$\pm$1  &   0.05  \\
 \enddata
 
Notes: \\
a: Derived based on the DL07 model.\\
b: Derived based on MBB.\\
c: The uncertainties of the fractions as derived by the SED decomposition using DECOMPIR, are $\geq$10\%. 
\end{deluxetable}

\begin{deluxetable}{lcccc}

\tabletypesize{\normalsize}
\tablewidth{0pc}
\tablecaption{Scaling relations between \lir\ and various MIRI broad-band luminosities. y=$\alpha$+$\beta~\times$ x, where y = $logL_{\rm IR}$ and x the corresponding MIRI luminosities.}
\tablehead{

\colhead{y/x} &
\colhead{$log(L_{\rm 10}/L_{\odot})$} &
\colhead{$log(L_{\rm 18}/L_{\odot})$} &
\colhead{$log(L_{\rm 21}/L_{\odot})$} &
\colhead{$log(L_{\rm 25.5}/L_{\odot})$}}

\startdata
 $log(L_{\rm IR}/L_{\odot})$ (SF) & $\alpha =$ 1.33$\pm$0.32, $\beta =$ 0.98$\pm$0.03   & 1.52$\pm$0.22, 0.95$\pm$0.02 & 1.98$\pm$0.27, 0.90$\pm$0.03  & 1.69$\pm$0.28, 0.91$\pm$0.03  \\
 
 $log(L_{\rm IR}/L_{\odot})$ (AGN) & $\alpha =$ 2.34$\pm$0.72, $\beta =$ 0.84$\pm$0.06   & 1.55$\pm$0.65, 0.90$\pm$0.06 & 1.46$\pm$0.64, 0.91$\pm$0.06  & 1.39$\pm$0.69, 0.91$\pm$0.06

\enddata
\end{deluxetable}

\end{document}